\documentclass[fleqn,usenatbib]{mnras}
\usepackage{newtxtext,newtxmath}
\usepackage[T1]{fontenc}
\pdfoutput=1    

\DeclareRobustCommand{\VAN}[3]{#2}
\let\VANthebibliography\thebibliography
\def\thebibliography{\DeclareRobustCommand{\VAN}[3]{##3}\VANthebibliography}

\usepackage{xspace}
\usepackage{graphicx}
\usepackage{xcolor}

\newcommand{\camb}{\textsc{CAMB}\xspace}
\newcommand{\halofit}{\textsc{HALOFIT}\xspace}
\newcommand{\mcal}{\textsc{metacalibration}\xspace}
\newcommand{\im}{\textsc{im3shape}\xspace}
\newcommand{\cosmolike}{\textsc{CosmoLike}\xspace}
\newcommand{\flask}{\textsc{flask}\xspace}
\newcommand{\namaster}{\textsc{NaMASTER}\xspace}
\newcommand{\healpix}{\textsc{HEALPix}\xspace}
\newcommand{\multinest}{\textsc{Multinest}\xspace}
\newcommand{\cosmosis}{\textsc{CosmoSIS}\xspace}

\title[DES-Y1 Cosmic Shear in Harmonic Space]{%
  Cosmic Shear in Harmonic Space from the Dark Energy Survey Year 1 Data:
  Compatibility with Configuration Space Results
}

\author[DES Collaboration]{
\parbox{\textwidth}{
\Large
H.~Camacho,$^{1,2}$\thanks{E-mail: hocamachoc@gmail.com (HC)}
F.~Andrade-Oliveira,$^{1,2}$
A.~Troja,$^{3,2}$
R.~Rosenfeld,$^{3,2}$\thanks{E-mail: rogerio.rosenfeld@gmail.com (RR)}
L.~Faga,$^{4,2}$
R.~Gomes,$^{4,2}$
C.~Doux,$^{5}$
X.~Fang,$^{6,7}$
M.~Lima,$^{4,2}$
V.~Miranda,$^{7}$
T.~F.~Eifler,$^{7,8}$
O.~Friedrich,$^{9}$
M.~Gatti,$^{5}$
G.~M.~Bernstein,$^{5}$
J.~Blazek,$^{10,11}$
S.~L.~Bridle,$^{12}$
A.~Choi,$^{13}$
C.~Davis,$^{14}$
J.~DeRose,$^{15}$
E.~Gaztanaga,$^{16,17}$
D.~Gruen,$^{18}$
W.~G.~Hartley,$^{19}$
B.~Hoyle,$^{18}$
M.~Jarvis,$^{5}$
N.~MacCrann,$^{20}$
J.~Prat,$^{21,22}$
M.~M.~Rau,$^{23}$
S.~Samuroff,$^{23}$
C.~S{\'a}nchez,$^{5}$
E.~Sheldon,$^{24}$
M.~A.~Troxel,$^{25}$
P.~Vielzeuf,$^{26}$
J.~Zuntz,$^{27}$
T.~M.~C.~Abbott,$^{28}$
M.~Aguena,$^{2}$
S.~Allam,$^{29}$
J.~Annis,$^{29}$
D.~Bacon,$^{30}$
E.~Bertin,$^{31,32}$
D.~Brooks,$^{33}$
D.~L.~Burke,$^{14,34}$
A.~Carnero~Rosell,$^{2}$
M.~Carrasco~Kind,$^{35,36}$
J.~Carretero,$^{26}$
F.~J.~Castander,$^{16,17}$
R.~Cawthon,$^{37}$
M.~Costanzi,$^{38,39,40}$
L.~N.~da Costa,$^{2,41}$
M.~E.~S.~Pereira,$^{42,43}$
J.~De~Vicente,$^{44}$
S.~Desai,$^{45}$
H.~T.~Diehl,$^{29}$
P.~Doel,$^{33}$
S.~Everett,$^{46}$
A.~E.~Evrard,$^{47,42}$
I.~Ferrero,$^{48}$
B.~Flaugher,$^{29}$
P.~Fosalba,$^{16,17}$
D.~Friedel,$^{35}$
J.~Frieman,$^{29,22}$
J.~Garc\'ia-Bellido,$^{49}$
D.~W.~Gerdes,$^{47,42}$
R.~A.~Gruendl,$^{35,36}$
J.~Gschwend,$^{2,41}$
G.~Gutierrez,$^{29}$
S.~R.~Hinton,$^{50}$
D.~L.~Hollowood,$^{46}$
K.~Honscheid,$^{51,52}$
D.~Huterer,$^{42}$
D.~J.~James,$^{53}$
K.~Kuehn,$^{54,55}$
N.~Kuropatkin,$^{29}$
O.~Lahav,$^{33}$
M.~A.~G.~Maia,$^{2,41}$
J.~L.~Marshall,$^{56}$
P.~Melchior,$^{57}$
F.~Menanteau,$^{35,36}$
R.~Miquel,$^{58,26}$
R.~Morgan,$^{59}$
F.~Paz-Chinch\'{o}n,$^{35,60}$
D.~Petravick,$^{35}$
A.~Pieres,$^{2,41}$
A.~A.~Plazas~Malag\'on,$^{57}$
K.~Reil,$^{34}$
M.~Rodriguez-Monroy,$^{44}$
E.~Sanchez,$^{44}$
V.~Scarpine,$^{29}$
M.~Schubnell,$^{42}$
S.~Serrano,$^{16,17}$
I.~Sevilla-Noarbe,$^{44}$
M.~Smith,$^{61}$
M.~Soares-Santos,$^{42}$
E.~Suchyta,$^{62}$
G.~Tarle,$^{42}$
D.~Thomas,$^{30}$
C.~To,$^{63,14,34}$
T.~N.~Varga,$^{64,65}$
J.~Weller,$^{64,65}$
and R.D.~Wilkinson$^{66}$
\begin{center} (DES Collaboration) \end{center}
}
}

\date{Accepted 2022 August 28. Received 2022 August 2; in original form 2021 November 28}
\pubyear{2022}

\begin{document}
\label{firstpage}
\pagerange{\pageref{firstpage}--\pageref{lastpage}}
\maketitle

\begin{abstract}
  We perform a cosmic shear analysis in harmonic space using the first year of
  data collected by the Dark Energy Survey (DES-Y1). We measure the cosmic weak
  lensing shear power spectra using the \mcal catalogue and perform a
  likelihood analysis within the framework of CosmoSIS. We set scale cuts based
  on baryonic effects contamination and model redshift and shear calibration
  uncertainties as well as intrinsic alignments. We adopt as fiducial
  covariance matrix an analytical computation accounting for the mask geometry
  in the Gaussian term, including non-Gaussian contributions. A suite of 1200
  lognormal simulations is used to validate the harmonic space pipeline and the
  covariance matrix. We perform a series of stress tests to gauge the
  robustness of the harmonic space analysis.
  Finally, we use the DES-Y1 pipeline in configuration space to perform a similar likelihood analysis and compare both results. demonstrating their compatibility in estimating the cosmological parameters $S_8$, $\sigma_8$ and $\Omega_m$.
  We use the DES-Y1 \mcal shape catalogue, with photometric redshifts estimates in the range $0.2-1.3$, divided in four tomographic bins finding $\sigma_8(\Omega_m/0.3)^{0.5} = 0.766\pm 0.033$ at 68\% CL.
  The methods implemented and validated in this paper will allow us to perform a consistent harmonic space analysis in the upcoming DES data.
\end{abstract}



\section{Introduction}
\label{sec:introduction}
One of the consequences of the Theory of General Relativity is the precise prediction of the deflection of light due to the presence of matter in its path \citep{Einstein:1916vd}. This prediction was confirmed for the first time with the measurements of the positions of stars during a solar eclipse in 1919  by two expeditions, sent to Brazil and to the Principe Island \citep{Dyson:1920cwa}.
After roughly 100 years, and the enormous development of instrumental and theoretical methods, one is able to measure minute distortions in the shape of distant galaxies 
that provide information about the distribution of matter in the universe. These small distortions are called weak gravitational lensing, in opposition to strong gravitational lensing, when large distortions with multiple images of the same object are produced (for reviews see, e.g. \citet{Bartelmann:1999yn,2015RPPh...78h6901K,2017grle.book.....D,2018ARA&A..56..393M}).

Being a small effect, weak gravitational lensing can be detected only by capturing the images of a large sample of galaxies, usually called source galaxies, and performing shape measurements that can then be analyzed statistically. 
One of the most common ways to analyse weak lensing signals is by studying the correlation between shapes of two galaxies. This can be done in configuration space, with measurements of the two-point correlation functions, or in harmonic space and the corresponding measurement of the power spectra. Although 
they are both second order statistics and can be related by a Fourier transform, they probe scales differently, and so they behave differently to systematic effects and analysis choices. In practice, 
there are differences in the measurements and analyses that may yield different cosmological results from the configuration and harmonic space methods \citep{Hamana:2019etx}. 
In particular, the covariance matrix is known to be more diagonal (indicating less cross-correlations) in harmonic space than in configuration space due to the orthogonality of the spherical harmonics used to decompose the signal 
(see e.g., figure 2 in~\cite{y3-BAOkp}).
The consistency between cosmic shear analyses in configuration and harmonic space was recently investigated in \citet{Doux:2020duk},
using DES-Y3-like Gaussian mock catalogues and paying particular attention to the methodology of determining angular and multipole scale cuts in both cases.

In the past years, several collaborations reported results from weak gravitational lensing: the Deep Lens Survey (DLS)\footnote{\tt dls.physics.ucdavis.edu}, the Canada-France-Hawaii Telescope Lensing Survey (CFHTLenS)\footnote{\tt www.cfhtlens.org},  the Hyper Suprime-Cam Subaru Strategic Program (HSC-SSP)\footnote{\tt hsc.mtk.nao.ac.jp/ssp}, the Kilo-Degree Survey (KiDS)\footnote{\tt kids.strw.leidenuniv.nl} and the Dark Energy Survey (DES)\footnote{\tt www.darkenergysurvey.org}. 
DLS \citep{Jee:2012hr,Jee:2015jta} and CFHTLenS \citep{Joudaki:2016mvz} presented results from configuration space measurements whereas
HSC has performed the analysis both in harmonic space \citep{Hikage:2018qbn} and configuration space \citep{Hamana:2019etx}. KiDS has performed a cosmic shear analysis in configuration space for its 450 deg$^2$ survey \citep{Hildebrandt:2016iqg} and for its fourth  data release (KiDS-1000) a first comparison of configuration and harmonic space analyses was presented in \citep{Asgari:2020wuj} using bandpowers constructed from correlation functions, and more recently in \citep{2021arXiv211006947L} using the angular power spectrum forward modelling survey geometry effects, both showed excellent agreement.
For its first year of data (Y1), DES has presented a weak lensing analysis in configuration space only \citep{troxel_2018}. 

Two re-analyses of DES-Y1 weak gravitational lensing in combination with other experiments have been performed: KiDS-450 and DES-Y1 \citep{Joudaki:2019pmv}, and  DLS, CFHTLens, KiDS-450 and the DES Science Verification data \citep{Chang:2018rxd}. More recently, the DES-Y1 public data was used to perform a full 3x2pt analysis (the combination of shear, galaxy clustering and galaxy-galaxy lensing) in harmonic space with emphasis on the testing of a more sophisticated model for galaxy bias \citep{hadzhiyska2021hefty}. 

Consistency between different summary statistics analyses is expected when applied to the same data set.
As different statistics summarize information differently and could be sensitive to different systematic effects, consistency not only adds to the robustness of the different analyses and data reduction but also prevents ambiguity when comparing different data sets or analyses.
Nevertheless, recent studies have presented some tension on recovered parameters at the 0.5 to 1.5$\sigma$ between configuration and harmonic space analysis on the same data set.
See, e.g., cosmic shear analysis from the HSC \citep{Hamana:2019etx,Hikage:2018qbn,2022PASJ...74..488H}.
These tensions, although somehow small and understood in terms of the different scales probed, deserve consideration and showcase the importance of running both analyses in parallel for forthcoming galaxy surveys to understand better the capabilities and limitations of different two-point statistics.

The purpose of this paper is to complete the Y1 weak lensing analysis by presenting harmonic space results and comparing them to the configuration space ones. We measure the cosmic weak lensing shear power spectra  using the so-called \mcal catalogs\citep{mcal1,mcal2,zuntz:2017pso}. 
We perform a likelihood
analysis using the framework of CosmoSIS adopted by DES \citep{troxel_2018} assuming a fiducial $\Lambda$CDM cosmological model with parameters given in the Table \ref{tab:Parameters}. We use 1200 lognormal simulations originally developed for DES-Y1 \citep{krause:2017ekm} to validate
an analytical covariance matrix and scale cuts tested to curb the contributions from baryonic effects to the
shear power spectra. 
To demonstrate the compatibility between our analysis in harmonic space with the DES default analysis in configuration space, we run the DES-Y1 standard configuration space pipeline with a similar likelihood analysis methodology.
One of the main consequences of this work is to put forward a harmonic space analysis of galaxy shear validated with DES-Y1 data that justifies its adoption in an independent harmonic analysis with the DES-Y3 data \citep{2022MNRAS.tmp.1761D} and in the current analyses of the final six-years data set.

This paper is organized as follows.
Section~\ref{sec:formalism} reviews the basic theoretical modelling, including systematic effects such as redshift uncertainties, shear calibration and intrinsic alignments. In Section~\ref{sec:data} we describe the DES-Y1 data for the shear analysis presented here, Section~\ref{sec:Flask} presents the 1200 \flask lognormal mocks used to validate our pipeline and the analytical covariance matrix. Section~\ref{sec:methods} details our methodology including a discussion of the covariance matrix.
We perform likelihood analyses both in harmonic and configuration space and present our main results in Section~\ref{sec:results}, with some robustness tests shown in Section~\ref{sec:robustness}. We conclude in Section~\ref{sec:conclusions}.

\begin{table}
  \caption{The cosmological and nuisance parameters used in Y1 analysis. The
    fiducial values were used in the generation of the 1200 \flask mocks for
  DES-Y1. The priors were used for DES-Y1 real-space likelihood analysis.}
  \label{tab:Parameters}
  \begin{tabular*}{\columnwidth}{@{}l@{\hspace*{45pt}}c@{\hspace*{45pt}}c@{}}
    \hline
    Parameter & Fiducial value & Prior \\
    \hline
    $\Omega_m$ & 0.286 & $U(0.1, 0.9)$ \\
    $h$ & 0.70 & $U(0.55, 0.90)$ \\
    $\Omega_b$ & 0.05 & $U(0.03, 0.07)$ \\
    $n_s$ & 0.96 & $U(0.87, 1.07)$ \\
    $A_s\times 10^{9}$ & 2.232746 & $U(0.5, 5.0)$ \\
    $\Omega_{\nu}h^2$ & 0.0 & $U(0.0, 0.01)$ \\
                      & & \\
    $A_{\rm IA}$ & 0 & $U(-5.0, 5.0)$ \\
    $\alpha_{\rm IA}$ & 0 & $U(-5.0, 5.0)$ \\
                      & & \\
    ($m^1$ -- $ m^4$)$\times 10^2$ & 0 & $N(1.2, 2.3)$ \\
    $\Delta z^1\times10^2$ & 0 & $N(-0.1, 1.6)$ \\
    $\Delta z^2\times10^2$ & 0 & $N(-1.9,1.3)$ \\
    $\Delta z^3\times10^2$ & 0 & $N(0.9, 1.1)$ \\
    $\Delta z^4\times10^2$ & 0 & $N(-1.8, 2.2)$ \\
    \hline
  \end{tabular*}
\end{table}


\section{Theoretical modelling}
\label{sec:formalism}

The distortion of the shape of an object due to the intervening matter is
described by a lensing potential $\varphi(\vec{\theta})$ that is related to the
projection of the gravitational potential $\Phi(\vec{r})$ along the
line-of-sight from the source (S) to us (we will denote the comoving distance
by $\chi$ and use units where $c=1$):
\begin{equation}
  \varphi(\vec{\theta}) = \frac{2}{\chi_S} \int_0^{\chi_S} d\chi \,
  \frac{\chi_S-\chi}{\chi} \Phi(\chi,\vec{\theta}).
\end{equation}

The convergence ($\kappa$) and shear ($\gamma_1$ and $\gamma_2$) fields are
derived from the lensing potential, $\varphi$ as~\footnote{Following~\citet{troxel_2018},
throughout this work we assume the flat-sky approximation.}:
\begin{eqnarray}
  \kappa(\vec{\theta}) &=&  \frac{1}{2} \left(\frac{\partial^2
    \varphi}{\partial \theta_1^2} + \frac{\partial^2 \varphi}{\partial
    \theta_2^2} \right),  \\
  \gamma_1(\vec{\theta}) &=& \frac{1}{2} \left(\frac{\partial^2
    \varphi}{\partial \theta_1^2} - \frac{\partial^2 \varphi}{\partial
    \theta_2^2} \right),  \\
  \gamma_2(\vec{\theta}) &=& \frac{\partial^2 \varphi}{\partial \theta_1
    \partial \theta_2} ;
\end{eqnarray}
where $\theta_{1,2}$ are the sky coordinates.

Using the Poisson equation one can write the convergence in terms of the
density perturbation $\delta = \delta \rho/\bar{\rho}$ as:
\begin{equation}
  \kappa(\vec{\theta}) = \int_0^{\chi_S} d\chi \, W_\kappa(\chi)
  \delta(\chi,\vec{\theta}),
\end{equation}
where the lensing window function $W_\kappa(\chi)$ can be defined by:
\begin{equation}
  W_\kappa(\chi) = \frac{3 H_0^2 \Omega_m \chi}{2 a(\chi)} \int_\chi^{\chi_H}
  d\chi_S \; \frac{dn}{dz}(z(\chi_S)) \frac{dz}{d \chi_S} \left( 1-
  \frac{\chi}{\chi_S} \right),
\end{equation}
where $\chi_H$ is the comoving distance to the cosmic horizon, $H_0$ the Hubble constant, $\Omega_m$ the matter density parameter, $a(\chi)$ the scale factor and for multiple
galaxy sources described by a redshift distribution normalised as:
\begin{equation}
  \int_0^\infty dz \; \frac{dn}{dz}(z) = 1,
\end{equation}
with $\frac{dn}{dz}$ the redshift distribution of galaxies.
Note here we're assuming a flat $\Lambda$CDM cosmological model.

In harmonic space we can write the convergence and shear fields as:
\begin{eqnarray}
  \kappa(\vec{\ell}) &=&  -\frac{|\ell|^2}{2} \varphi(\vec{\ell}),  \\
  \gamma_1(\vec{\ell}) &=& \frac{\ell_2^2 - \ell_1^2}{2} \varphi(\vec{\ell}),
  \\
  \gamma_2(\vec{\ell}) &=& -\ell_1 \ell_2 \varphi(\vec{\ell});
\end{eqnarray}
where $\ell_{1,2}$ are Fourier conjugated variables of $\theta_{1,2}$.

The convergence and shear fields are not independent since they are determined
by the gravitational potential. One can find linear combinations of $\gamma_1$
and $\gamma_2$, the so-called $E$ and $B$ modes denoted by $\gamma_E$ and
$\gamma_B$ such that:
\begin{equation}
  \gamma_E(\vec{\ell}) = \kappa(\vec{\ell}); \;\; \gamma_B(\vec{\ell}) =0.
\end{equation}

Finally, we are interested in the 2-point correlations between these fields. In
the Limber
approximation~\citep{1953ApJ...117..134L,1992ApJ...388..272K,2008PhRvD..78l3506L,2017MNRAS.469.2737K,2017JCAP...05..014L,2017MNRAS.472.2126K}
the $E$-mode angular power spectrum $C^{EE}(\ell)$ (which is equal to the
convergence angular power spectrum $C^{\kappa \kappa}(\ell) $) is given by:
\begin{equation}
  C^{EE}_{(i,j)}(\ell) = \int_0^{\chi_H} d\chi \; \frac{W^i_\kappa(\chi)
    W^j_\kappa(\chi)}{\chi^2}
  P_{m}\left(\frac{\ell+1/2}{\chi},z(\chi) \right),
\end{equation}
where we have introduced indices for the different tomographic redshift bins
$(i,j)$ that will be used in the analyses and $P_m$ is the total matter power
spectrum, modelled here to include nonlinear effects using the \camb Boltzmann
solver~\citep{2000ApJ...538..473L,2012JCAP...04..027H} and the
\halofit~\citep{2003MNRAS.341.1311S} prescription with updates from
\citet{2012ApJ...761..152T}. The shear angular correlation functions
$\xi_{\pm}(\theta)$ that are also used in the comparison performed in this
paper can be computed from the angular power spectra (see, {\it e.g.} equation
(9) in \citet{2020arXiv201208568F}).

We also model three astrophysical and observational systematic effects using
the DES-Y1 methodology (see details in \citet{krause:2017ekm,troxel_2018}):
\begin{itemize}
  \item{Redshift distributions:} an additive bias $\Delta z^i$ on the mean of
    the redshift distribution of source galaxies in each tomographic bin $i$ is
    introduced to account for uncertainties on the photometric redshift
    estimation;
  \item{Shear calibration:} a multiplicative bias on the shear amplitude is
    included in each tomographic bin $i$ to account for uncertainties on the
    shear calibration and included in our power spectra modelling
    as~\citep{2006MNRAS.368.1323H,2006MNRAS.366..101H}  ;
  \item{Intrinsic alignments:} we use the nonlinear alignment model
    (NLA)~\citep{2012MNRAS.424.1647K,2007NJPh....9..444B} for the intrinsic
    alignment corrections to the cosmic-shear power spectrum.
    Our model for the observed cosmic shear $EE$ power spectra is given by
    $C_{i,j}(\ell) = C_{i,j}^{\rm GG}(\ell) + C_{i,j}^{\rm GI}(\ell) + C_{i,j}^{\rm IG}(\ell) + C_{i,j}^{\rm II}(\ell)$,
    where `G and `I' stands for `Gravitational' and `Intrinsic' shear signals, so that the `GG' term refers to the pure cosmic shear signal.
    The remaining terms accounts for its correlations with galaxy intrinsic alignments.
    See \citep{troxel_2018,krause:2017ekm} for further details on the DES-Y1 IA modeling and \citep{2015PhR...558....1T,2015SSRv..193....1J} for general IA effect reviews.
    The amplitude of those terms is scaled as $C^{GI,IG} \propto A$ and $C^{II} \propto A^2$ by a nonlinear alignment amplitude, $A$, with a redshift dependence parametrised as $A = A_{\rm IA} \left[ (1+z) / (1+z_0) \right]^{\alpha_{\rm IA}}$, with $z_0 = 0.62$ fixed at approximately
    the mean redshift of source galaxies and $A_{\rm IA}$,
    $\alpha_{\rm IA}$ are free parameters in our model\footnote{In the DES-Y1
    analysis a more sophisticated `tidal alignment and tidal torquing' (TATT)
    model \citep{TATT} for intrinsic alignment was also considered and found to
    be not required for the Y1 configuration. It became the fiducial choice in
    DES-Y3.}.
\end{itemize}

All the different pieces for the modelling presented above are used as modules
in the, publicly available, \cosmosis framework \citep{zuntz:2014csq}, in an
analogous way to what was done for the configuration-space analysis presented
in~\cite{troxel_2018}.
Finally, the theoretical angular power spectrum is binned 
into bandpowers.
This is done by filtering the predictions with a set of bandpower windows, $\mathcal{F}_{q\ell}^{ab}$, consistent with the pseudo-$C_\ell$ approach we follow for the data estimates (see section~\ref{sec:Measurements}).
Thus the final model for a bandpower, $\ell \in q$, is computed as
\begin{equation}
\label{eq:bpws-modeling}
    \mathbf{C}_{(i,j)}(q) = \sum_{\ell \in q} \mathcal{F}_{q \ell}^{(i,j)} \mathbf{C}_{(i,j)}(\ell)
\end{equation}
where $(i, j)$ represents the tomographic redshift bin pair, and a vector notation is required, $\mathbf{C} = \left( C^{EE}, C^{EB}, C^{BB} \right)$, to account for the $E-B$ mode decomposition of the shear field.
We refer the reader to \citet{2019MNRAS.484.4127A} for the somehow lengthy expressions for the bandpower windows and details about the $E-B$ mode decomposition.
The data, priors and redshift distributions are introduced in the following section.

\begin{table}
  \caption{Effective angular number density and shear dispersion for each
  tomographic redshift bin.}
  \label{tab:ShapeNoise}
  \begin{tabular*}{\columnwidth}{@{}l@{\hspace*{50pt}}c@{\hspace*{50pt}}c@{}}
    \hline
    redshift bin & $n_\mathrm{eff}$ & $\sigma_{\mathrm{e}}$ \\
    \hline
    $0.20 < z_\mathrm{phot} < 0.43$ & 1.5 & 0.3 \\
    $0.43 < z_\mathrm{phot} < 0.63$ & 1.5 & 0.3 \\
    $0.63 < z_\mathrm{phot} < 0.90$ & 1.5 & 0.3 \\
    $0.90 < z_\mathrm{phot} < 1.30$ & 1.7 & 0.3 \\
    \hline
  \end{tabular*}
\end{table}

\begin{figure}
  \includegraphics[width=0.45\textwidth]{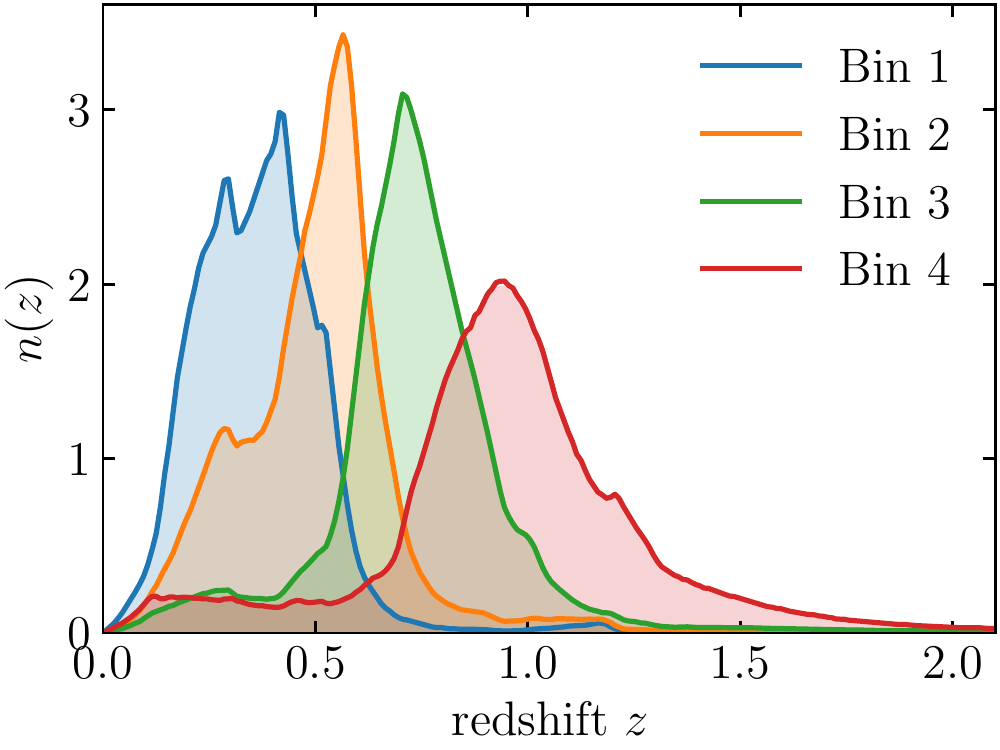}
  \caption{Redshift distributions for the four tomographic bins.
  See Table~\ref{tab:ShapeNoise}.}
  \label{fig:nz}
\end{figure}


\section{Data}
\label{sec:data}

The Dark Energy Survey (DES) conducted its six-year survey finalising in
January 2019 using a 570-megapixel camera mounted on the 4-meter Blanco
Telescope at the Cerro Tololo Inter-American Observatory (CTIO). The
photometric survey used five filters and collected information of more than 300
million galaxies in an area of roughly 5000 deg$^2$, allowing for the
measurement of shapes in addition to positions of galaxies.

The analysis of the first year of data \footnote{Public data products can be found in \url{https://des.ncsa.illinois.edu/releases/y1a1}}, denoted by DES-Y1, used two independent
pipelines \citep{zuntz:2017pso} to produce shape catalogues for its shear
analysis: \mcal \citep{mcal1,mcal2} and \im \citep{2013MNRAS.434.1604Z}. Here we will focus on the \mcal catalogue that was
used in the real-space fiducial analysis, with a final contiguous area of 1321
deg$^2$ containing 26 million galaxies with a density of 5.5 galaxies
arcmin$^{-2}$. 
A Bayesian Photometric Redshift (BPZ) \citep{BPZ} method was used to divide these source objects into four tomographic redshift bins shown in 
Table~\ref{tab:ShapeNoise}
with redshift distributions shown in Figure \ref{fig:nz}.
The priors on the redshift \citep{hoyle2018,davis,gatti2018} and shear calibration \citep{zuntz:2017pso} parameters are shown in the Table~\ref{tab:Parameters}.

In order to correct noise, modelling, and selection biases in the shear
estimate, one uses the \mcal method \citep{mcal1,mcal2}. It introduces a shear
response correction (a $2 \times 2$ matrix $R_i$ for each object $i$) that is
obtained by artificially shearing each image in the catalogue and has two
components: a response of the shape estimator and a response of the selection
of the objects. The DES Y1 \mcal catalogue does not implement any per-galaxy weight and the shear response
corrections are made available in the catalogue release\footnote{We note the improved DES Y3 \mcal catalogue now implements a per-galaxy weighting scheme, see \citep{gatti2021}}. The shear response is
used to obtain the estimated calibrated shear $\hat{\vec{\gamma}}_i$ for each
object from the measured ellipticities as \citep{zuntz:2017pso}:
\begin{equation}
    \hat{\vec{\gamma}}_i = \langle R_i \rangle^{-1} \vec{e}_i ,
    \label{eq:calibrated_shear}
\end{equation}
where we use an averaged response matrix for each tomographic redshift bin and
have also subtracted a nonzero mean $\left\langle \vec{e}_i \right\rangle $ per
tomographic bin prior to the shear estimation. The estimated shear per object
is pixelated in maps using the \healpix pixelisation scheme
\citep{Gorski:2004by} with a resolution $N_{side}=1024$ for each redshift
bin\footnote{All DES Y1/Y3 map-based analyses are performed at this resolution
because it is a good trade-off between resolution and number of galaxies per
pixel (see {\it e.g. }\citet{MassMapY1}). } and the angular power spectrum is
measured using \namaster \citep{2019MNRAS.484.4127A} as described in Section
\ref{sec:methods}.


\section{Lognormal mock catalogues}
\label{sec:Flask}

We use a set of 1200 lognormal realisations generated with the Full-sky
Lognormal Astro-fields Simulation Kit (\flask\footnote{\tt
www.astro.iag.usp.br/$\sim$flask}) \citep{Xavier:2016elr}, specially designed
for DES Y1 configuration-space analysis \citep{krause:2017ekm,troxel_2018} in
order to test our pipeline and validate the fiducial covariance presented in
this analysis.

The lognormal \flask realisations use as input the angular power spectrum for
each pair of redshift bins $(i,j)$. Those were computed using \cosmolike
\citep{krause:2016jvl} from a $\Lambda$CDM cosmological model with parameters
quoted as fiducial in Table~\ref{tab:Parameters} and redshift distributions for
four tomographic redshift bins that were used in the paper describing the
DES-Y1 methodology \citep{krause:2017ekm} and the paper reporting DES-Y1
cosmological results from cosmic shear \citep{troxel_2018}.

On top of the one- and two-point distributions, this suite of realisations were
also designed to match the reduced skewness of projected fields predicted by
perturbation theory at a fiducial scale of $10\, \mathrm{Mpc}\, h^{-1}$, see
\cite{2018PhRvD..98b3508F,krause:2017ekm} for details. This approach has been
shown to yield accurate results for DES-Y1 \citep{krause:2017ekm} and DES-Y3
\citep{2020arXiv201208568F} two-point observables. We also note that
\cite{2020arXiv201208568F} had shown, also in the context of DES analysis, that
the non-connected part of the covariance matrix does not cause significant bias
in a cosmological analysis.

The \flask shear maps are generated using \healpix with resolution set by an
$N_\mathrm{side}$ parameter of 4096. We further sample source galaxy positions
and ellipticity dispersion for each tomographic bin by matching the observed
number density of galaxies $n_{\rm eff}$ and the shape-noise parameter
$\sigma_{\rm e}$. The numbers used for the \flask mocks are given in
Table~\ref{tab:ShapeNoise} and are similar to the values used in
\citet{troxel_2018}.


\section{Methods}
\label{sec:methods}

\begin{figure*}
\centering \includegraphics[width=0.99\textwidth]{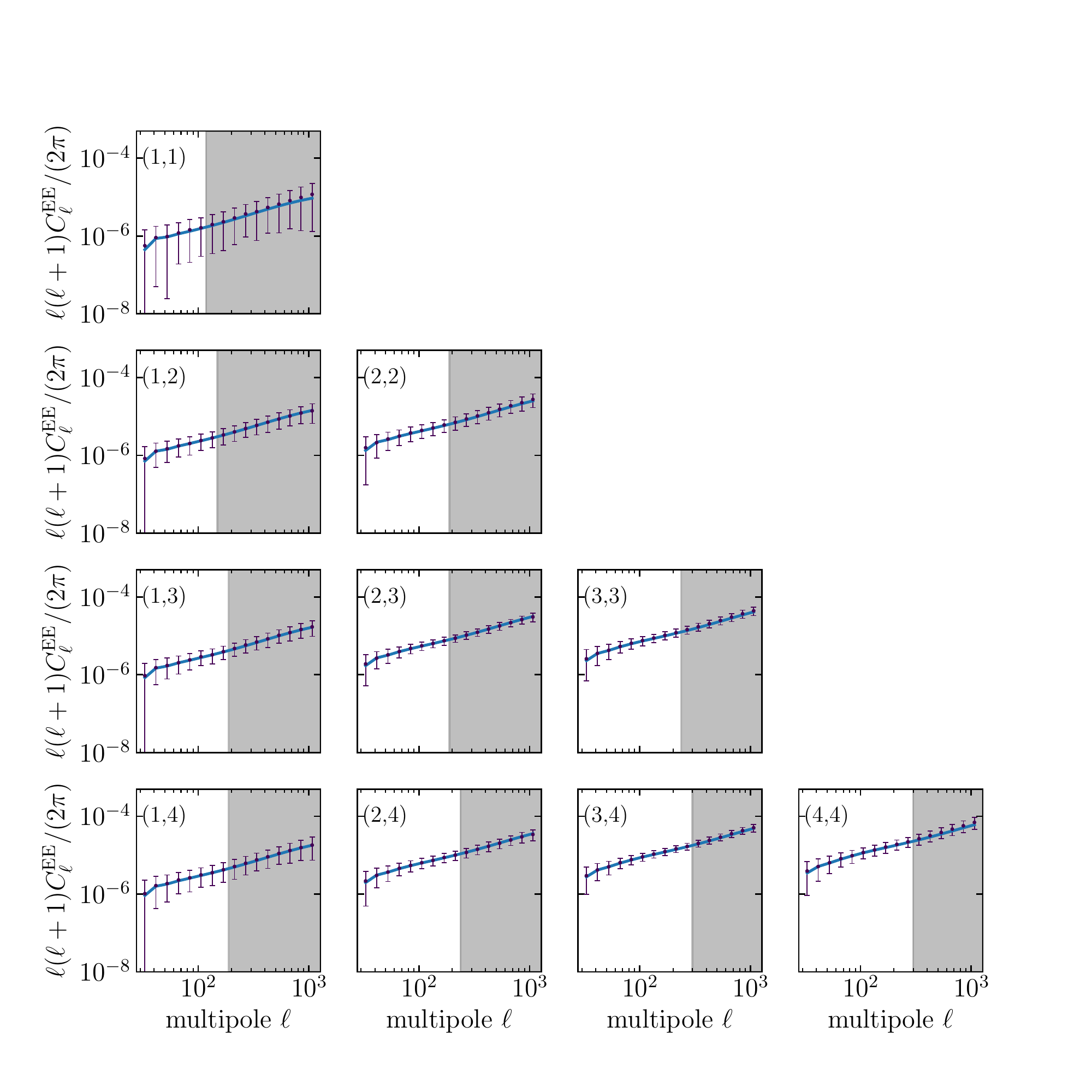}
\caption{Measured $C_\ell^{EE}$ cosmic shear angular power spectra on the 1200
  DESY1 \flask mocks. Points and error bars show the sample mean and standard
  deviation for the realizations. The continuous line is obtained from
  \cosmosis using the \flask cosmology and the vertical shaded regions shows
  the scale-cuts applied. }
\label{fig:flask_measurements}
\end{figure*}

\begin{figure}
  \includegraphics[width=0.50\textwidth]{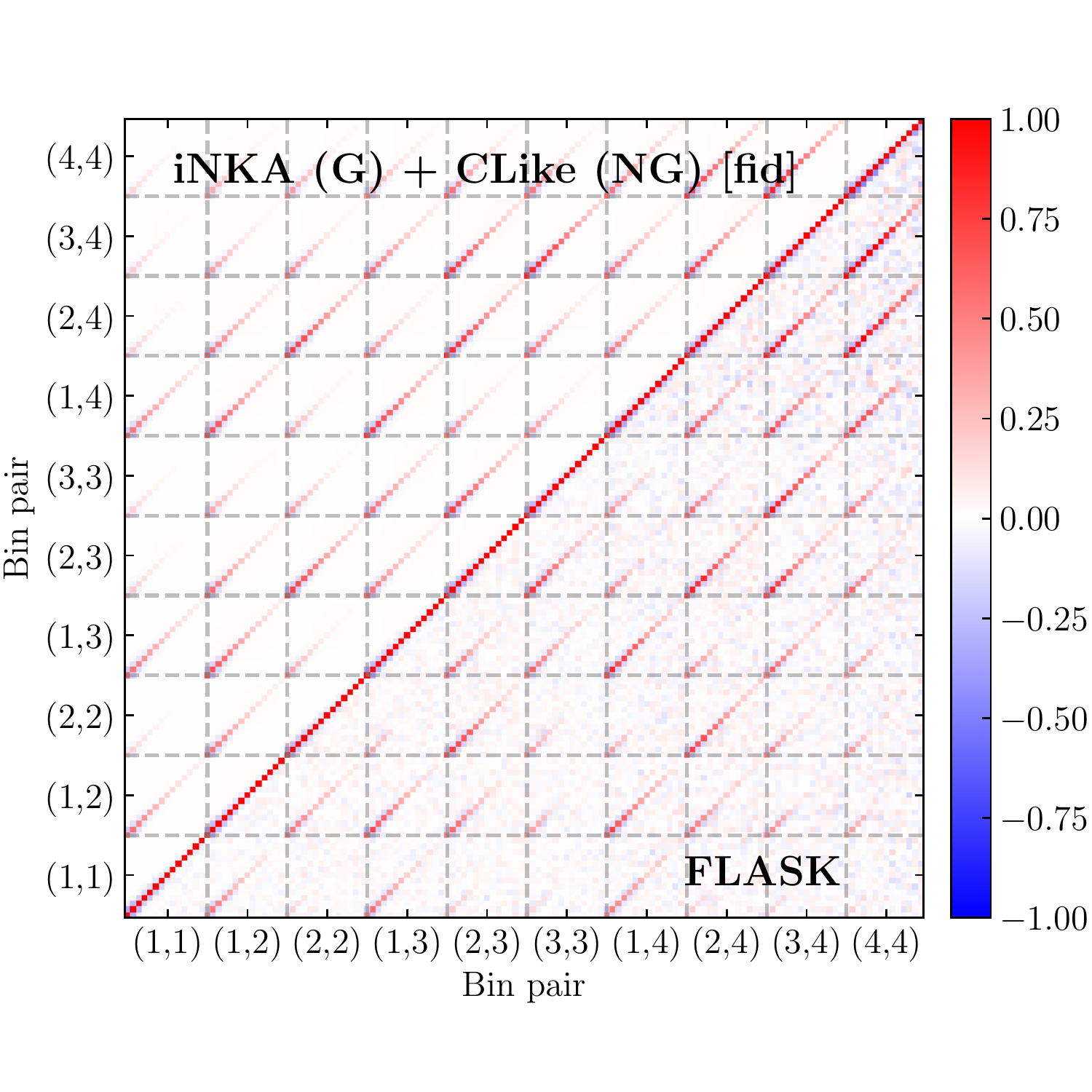}
  \caption{Comparison between our fiducial correlation matrix with the one
    obtained from the 1200 \flask mocks.
    We show the first 14 bandpower windows for readability and do not apply any scale cuts.
    }
  \label{fig:covariance}
\end{figure}

\begin{figure*}
  \includegraphics[width=\textwidth]{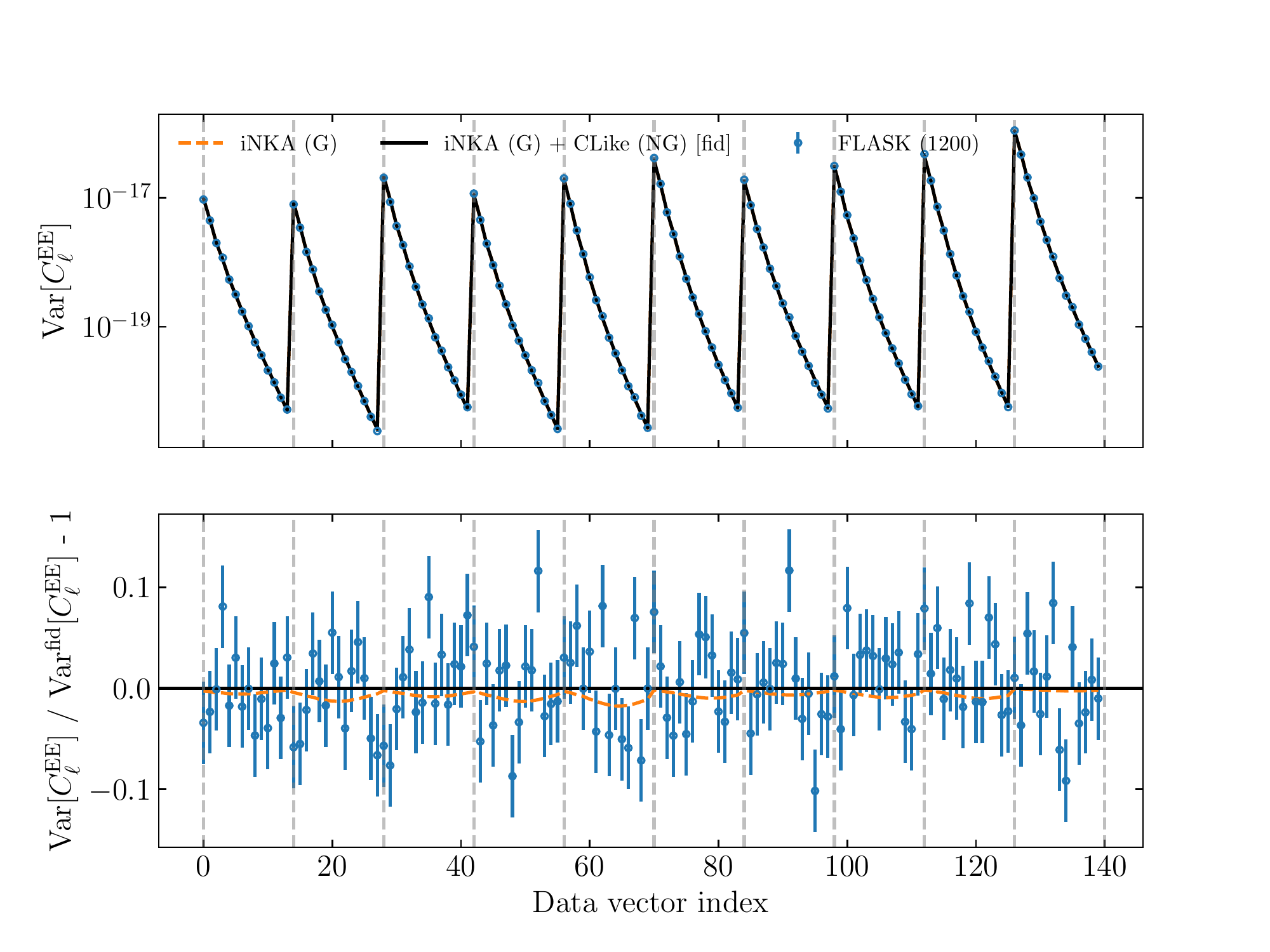}
  \caption{Diagonal elements of the different covariance matrices presented in this work.
      We show the first 14 bandpower windows for readability and do not apply any scale cuts.
      The top panel shows the absolute amplitude.
      Note that iNKA (G) and iNKA (G) + CLike (NG) [fid] lie almost one on top of the other.
      The bottom panel shows their relative difference.
      The vertical dashed lines represent the divisions between the different bin pairs considered, ordered as in Figure \ref{fig:covariance}.
    The DES-Y1 \flask mocks sample covariance error bars are computed using the Wishart distribution prediction ~\citep{2013MNRAS.432.1928T}.}
  \label{fig:variance}
\end{figure*}

\begin{table}
  \caption{Scale-cuts used for the fiducial analysis.
    The first column shows the tomographic bin pair and the second its scale cuts.
    We keep the large scale cut, smallest multipole considered, $\ell_{\rm min}$ fixed to $30$ and base our small scale cuts on a conservative one, based on the contribution from baryonic effects.
    Following \citep{troxel_2018}, we cut bandpowers with a fractional contribution grater than $2\%$ in our fiducial model.
  We use OWLS AGN simulation~\citep{2011MNRAS.415.3649V,2010MNRAS.402.1536S} to estimate this contribution.}
  \label{tab:scale-cuts}
  \begin{tabular*}{\columnwidth}{@{}l@{\hspace*{100pt}}c@{}}
    \hline
    Bin pair, $(a, b)$ & $[\ell_{\rm min}$, $\ell_{\rm max})$ \\
    \hline
    (1, 1) & [30, 150) \\
    (1, 2) & [30, 150) \\
    (1, 3) & [30, 189) \\
    (1, 4) & [30, 189) \\
    (2, 2) & [30, 238) \\
    (2, 3) & [30, 238) \\
    (2, 4) & [30, 189) \\
    (3, 3) & [30, 238) \\
    (3, 4) & [30, 300) \\
    (4, 4) & [30, 300) \\
    \hline
  \end{tabular*}
\end{table}

\begin{figure}
  \centering
  \includegraphics[width=0.49\textwidth]{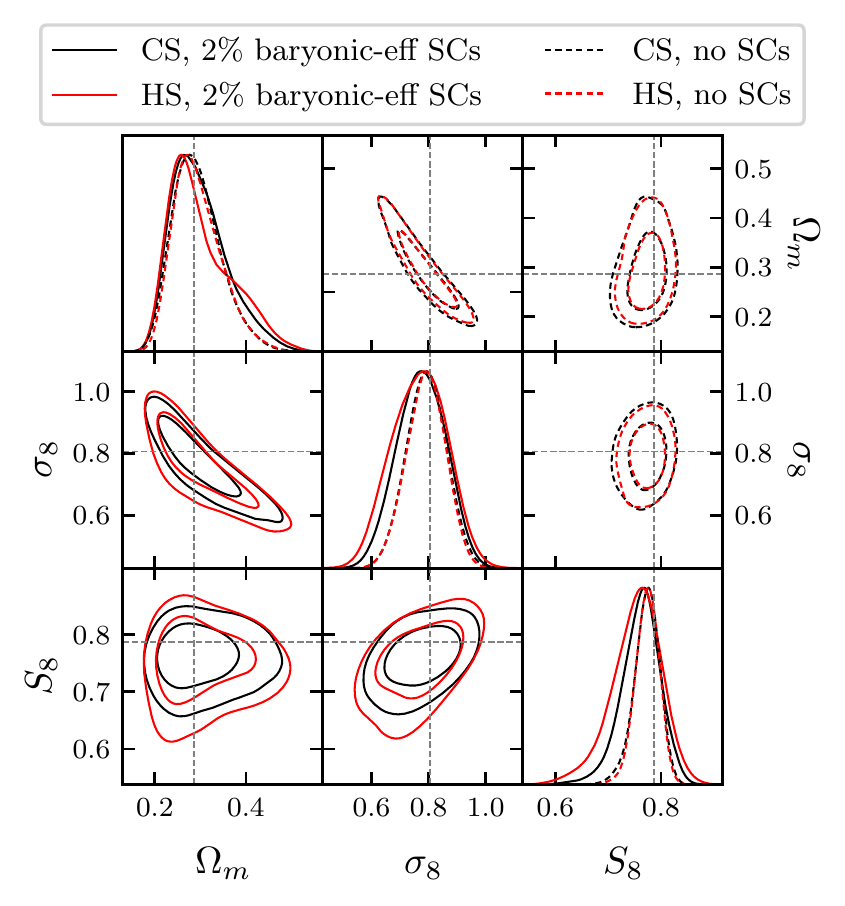}
  \caption{Marginalized posterior distributions for a subset of parameters for
    a noiseless data vector analyses in both configuration space (CS) and
    harmonic space (HS), with and without scale cuts (SC) Dashed lines are the
    input parameters.}
\label{fig:corner_hsvscs}
\end{figure}

In this section, we present the methodology to be used in our analysis. We
begin by describing the angular power spectra estimation, followed by a
discussion of the scale-cuts chosen to mitigate baryonic effects and end with a
discussion of the fiducial covariance matrix used in this work.

\subsection{Angular power spectrum measurements}
\label{sec:Measurements}

For the angular power spectra estimation, we use the so-called pseudo-$C_\ell$
or MASTER method~\citep{1973ApJ...185..413P,2005MNRAS.360.1262B,2002ApJ...567....2H}, as implemented in the \namaster
code~\footnote{\tt github.com/LSSTDESC/NaMaster}~\citep{2019MNRAS.484.4127A}.

For the pixelized representation of cosmic shear catalogs, we construct
weighted tomographic cosmic shear maps,
\begin{equation}
  \vec{\gamma}_p = \sum_{i\in p}v_i \hat{\vec{\gamma}}_i / \sum_{i\in p} v_i,
\end{equation}
where $p$ runs over pixels and $i\in p$ runs over the galaxies in each pixel,
$\hat{\vec{\gamma}}_i=(\hat\gamma_1, \hat\gamma_2)$ is the calibrated galaxy
shear (see eq.~(\ref{eq:calibrated_shear})) and $v_i$ its associated weight.
\footnote{As stated in section \ref{sec:data} the DES-Y1 \mcal catalogue do not implement any per-galaxy weighting scheme, thus $v_i = 1$ for all galaxies.}
Throughout this work we use a \healpix fiducial resolution
$N_\mathrm{side}=1024$, which corresponds to a typical pixel size of the order
of $3.4$ arcminutes.

In addition to the cosmic shear signal maps, the pseudo-$C_\ell$ method relies
on the use of an angular window function, also known as the mask. Such a mask
encodes the information of the partial-sky coverage of the observed signal and
is used to deconvolve this effect on the estimated bandpowers. In this work, we
use the sum of weights scheme presented in~\cite{Nicola:2020lhi}, and construct
tomographic mask maps as
\begin{equation}
  w_p = \sum_{i\in p}v_i\ ,
\end{equation}
where the $v_i$ are the individual galaxy weights assigned by \mcal. It is
important to notice that in this approach there are different masks constructed
for each tomographic bin, since the number of galaxies per pixel varies for
each bin. In practical terms, these masks are equivalent to the pixelised
weighted galaxy-count maps.

An important part of power spectra estimation is the so-called noise bias, always present on the raw signal auto-correlation measurements because of the discrete nature of the signal maps inherited from the galaxy catalogues, giving a Poissonian component. 
On top of that, for cosmic shear there is also a Gaussian component accounting for any systematic shape noise.
For the specific case of the pseudo-$C_\ell$ algorithm, the noise bias must be subtracted from the
auto-correlations in order to obtain an unbiased estimate of the signal power
spectrum. Schematically, the true binned power spectrum estimator can be
written as \citep{2019MNRAS.484.4127A}:
\begin{equation}
    \hat{C}^{ab}_q = \sum_{q'} ({\cal M}^{ab})^{-1}_{q q'}\left(
    \tilde{C}^{ab}_{q'} - \delta_{ab} \tilde{N}^b_{q'} \right),
\end{equation}
where $\delta_{ab}$ is the Kronecker delta, ${\cal M}^{ab}_{qq'} = \sum_{\ell\in q, \ell'\in q'} \mathsf{w}_q^\ell
\mathsf{M}^{ab}_{\ell\ell'}$ is the binned version of the coupling matrix,
$\mathsf{M}^{ab}_{\ell\ell'}$, that can be calculated analytically and depends
on the mask maps for the tomographic bins $a$ and $b$, $\tilde{C}^{ab}_q =
\sum_{\ell\in q} \mathsf{w}_q^\ell \tilde{C}^{ab}_\ell$ is
the binned version of the pseudo-$C_\ell$, $\tilde{C}^{ab}_\ell$. Here $q, q'$
represent multipole bins or bandpowers and $\mathsf{w}_q^\ell$ are multipole
weights defined for $\ell \in q$ and normalized to $\sum_{\ell\in q}
\mathsf{w}_q^{\ell} = 1$ \footnote{Troughout this work we assume equal weights
for all multipoles on each bandpower.}, see \citep{2019MNRAS.484.4127A} for
more details. Finally, $\tilde{N}_{q} = \sum_{\ell\in q} \mathsf{w}_q^\ell
\tilde{N}_\ell$ are the binned version of the noise bias pseudo-spectra,
$\tilde{N}_\ell$, given (in the sum of weights scheme) by
\citep{Nicola:2020lhi}:
\begin{equation}
  \label{eq:PCLNoise}
  \tilde{N}_\ell = A_{\rm pix} \left\langle \sum_{i\in p} v_i^2 \sigma_{\gamma,i}^2
  \right\rangle_{\rm pix},
\end{equation}
where the average $\left\langle \cdot \right\rangle_{\rm pix}$ is over all the
pixels, $A_{\rm pix}$ is the area of the pixels on the chosen \healpix
resolution and
\begin{equation}
  \sigma_{\gamma, i}^2 = \frac12 \left( \gamma_{1,i}^2 + \gamma_{2,i}^2 \right)
\end{equation}
is the estimated shear variance of each galaxy. Notice that the noise-bias
pseudo-power spectrum $\tilde{N_\ell}$ is independent of the $\ell$
multipole. The true noise-bias power spectrum $N_\ell$ is obtained using the
\namaster method, deconvolving the mask and performing the same $\ell$ binning
as the signal. This noise bias contribution, subtracted from the measurements,
must be included in the covariance matrix, as we will discuss below.

Finally, it is well known that the pixelization process of the shear field can introduce biases in its estimated pseudo-spectra.
We correct for the effect of pixelization by dividing the pseudo-spectra by the squared \healpix pixel window function $F_\ell$, i.e, $\tilde{C}^{ab}_\ell \to \tilde{C}^{ab}_\ell / F_\ell^2$.

\subsection{Binning and scale-cuts}
\label{sec:ScaleCuts}

\begin{figure}
  \centering
  \includegraphics[width=0.49\textwidth]{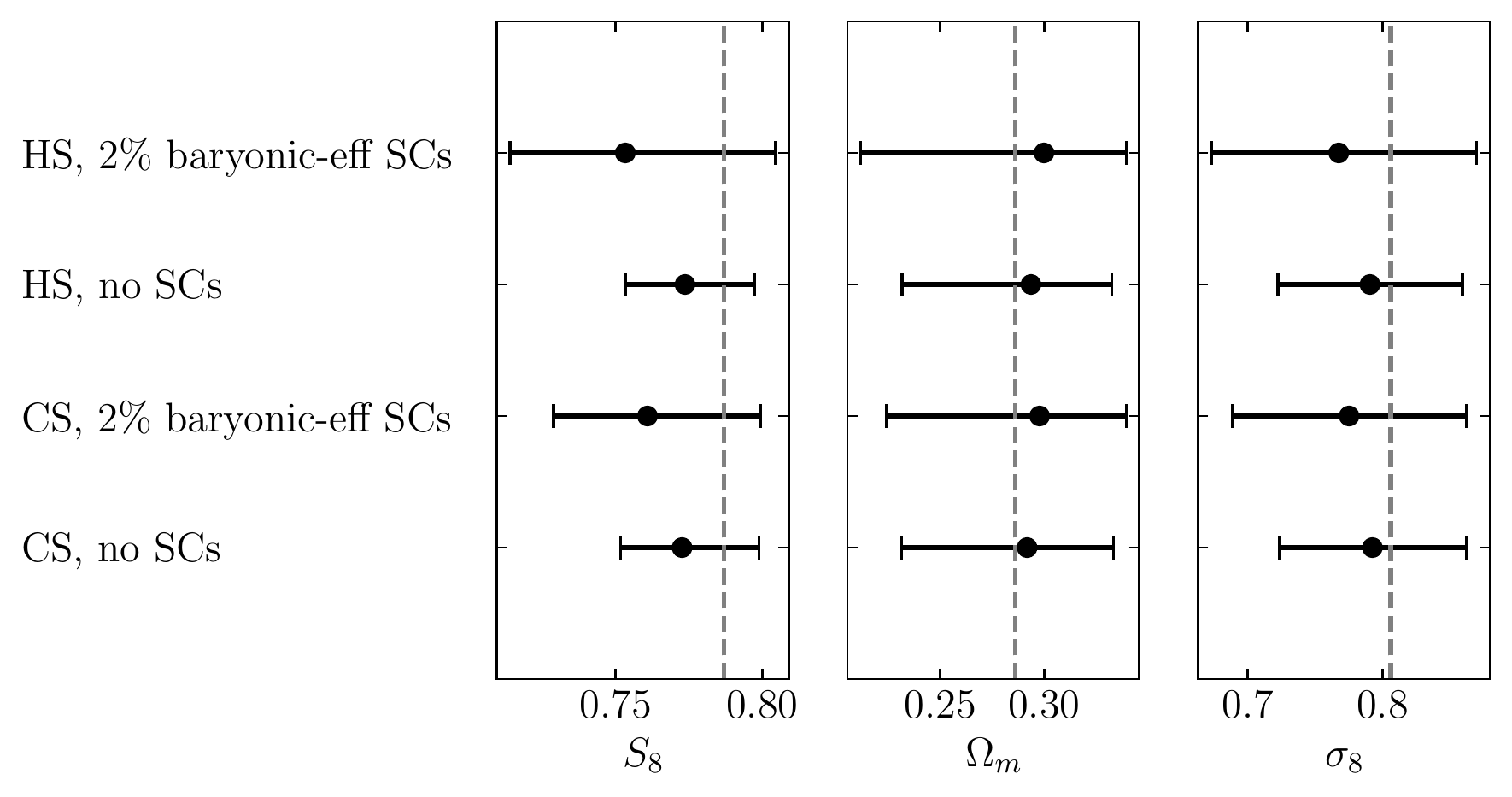}
  \caption{Marginalized constraints for the analyses from a noiseless data
    vector in harmonic space (HS) and configuration space (CS). We show
    results without scale cuts (``no SCs" in the figure) and with scale cuts
    motivated by baryonic effects (``$2\%$ baryonic-eff SCs" in the figure).}
\label{fig:robustness_flask}
\end{figure}

For all the angular power spectra measured here, we consider angular multipoles
$\ell \in$ [30, 3000) divided into $20$ logarithmic-spaced bandpowers with
edges similar to the binning scheme used in~\cite{2021arXiv210314190A}, where
analysis of DES-Y1 galaxy clustering in harmonic space is performed.

A comparison between the measured cosmic shear angular power spectra on the
mocks and the input theory prediction used for its generation is shown in
Figure \ref{fig:flask_measurements}. We find very good agreement, validating
the measurement pipeline (namely, the noise-subtraction method and the
computation of the coupling matrix).

Scale cuts are a key factor for cosmic shear analyses (see, {\it e.g.}
\citet{Doux:2020duk}). For the small scales, we follow the DES-Y1
configuration-space analysis and cut-out scales where baryonic effects
introduce a significant bias in the angular power spectra \citep{troxel_2018}.
To estimate the impact of baryon physics, the OWLS (OverWhelmingly Large
Simulations project) suite of hydrodynamic
simulations~\citep{2011MNRAS.415.3649V,2010MNRAS.402.1536S} is used for
re-scaling the computed non-linear power spectrum in our fiducial model
prediction by a factor
\begin{equation}
  P_{\textrm{NL}}(k) \rightarrow
  \frac{P_{\textrm{DM}+\textrm{Baryon}}(k)}{P_{\textrm{DM}}(k)} \times
  P_{\textrm{NL}}(k) ,
\end{equation}
where `DM' refers to the power spectrum from the OWLS dark-matter-only
simulation, while `DM+Baryon' refers to the power spectrum from the OWLS AGN
simulation~\citep{2011MNRAS.415.3649V,2010MNRAS.402.1536S}. It is important to
note that the particular use of the OWLS simulations, among others for DES-Y1
analysis, is a conservative choice, as they offer some of the most significant
deviations from the DM cases in the power spectrum \cite{troxel_2018}. We then
compare the predictions for the cosmic-shear angular power spectra with and
without the re-scaling for $P_{NL}(k)$ and impose, for our fiducial analysis,
the same $2\%$ threshold imposed by the configuration space analysis
\citep{troxel_2018}. Hence we remove from our data vector all bandpowers with
a fractional contribution from baryonic effects greater than $2\%$ in our
fiducial model for each pair of redshift bins.

We adopt a fiducial value for the lower multipole value $\ell \geq 30$ and test a
different value as a robustness test.
Our fiducial scale-cuts are summarized in table \ref{tab:scale-cuts}.
Our final data vector ends up having a
total of 85 entries. We note that an improvement should be expected by
including baryonic effects in the modelling and relaxing the proposed scale
cuts. As already shown in configuration space
\citep{2021MNRAS.502.6010H,2021arXiv210401397M} such improvement can be of
$\sim 20\%$ on the recovered constraints.

\subsection{Covariance matrix}
\label{sec:MethodsCovariance}

The covariance matrix has Gaussian, non-Gaussian and noise contributions and we
use two different methods to compute them. For the Gaussian contribution, we
rely on the so-called improved narrow-kernel approximation (iNKA) approach
within the pseudo-$C_{\ell}$ framework that takes into account the geometry of
the finite survey area described by the mask maps
\citep{Garcia-Garcia:2019bku,Nicola:2020lhi}.
We also use the full model for the noise terms in the pseudo-spectra Gaussian covariance as given by~\citet{Nicola:2020lhi} (their equation (2.29)).
The non-Gaussian contribution consists of the so-called super-sample covariance (SSC) and the
connected part of the 4-point function. These are obtained using the halo
model analytical computations with the \cosmolike code \citep{krause:2016jvl}
in harmonic space\footnote{DES-Y1 and Y3 analyses in configuration space use
\cosmolike covariance matrices as fiducial.}.

In order to validate our covariance model, we use measurements on the 1200
DES-Y1 \flask lognormal mocks to estimate a sample covariance matrix for the
angular power spectrum. In Figure~\ref{fig:covariance} we show a comparison
between our fiducial covariance matrix (computed at the \flask cosmology) and
the \flask covariance. One can see a good agreement, with the \flask covariance
being noisier in the non-diagonal elements, as expected.

A more quantitative comparison is presented in Figure~\ref{fig:variance}, where
we plot the diagonal elements of the two covariance matrices with error bars
obtained from a Wishart distribution. This figure shows that the contribution
from the non-Gaussian part to the diagonal of the covariance matrix is
negligible in our case.


\section{Likelihood analysis}
\label{sec:results}

\begin{figure*}
  \centering
  \includegraphics[width=0.99\textwidth]{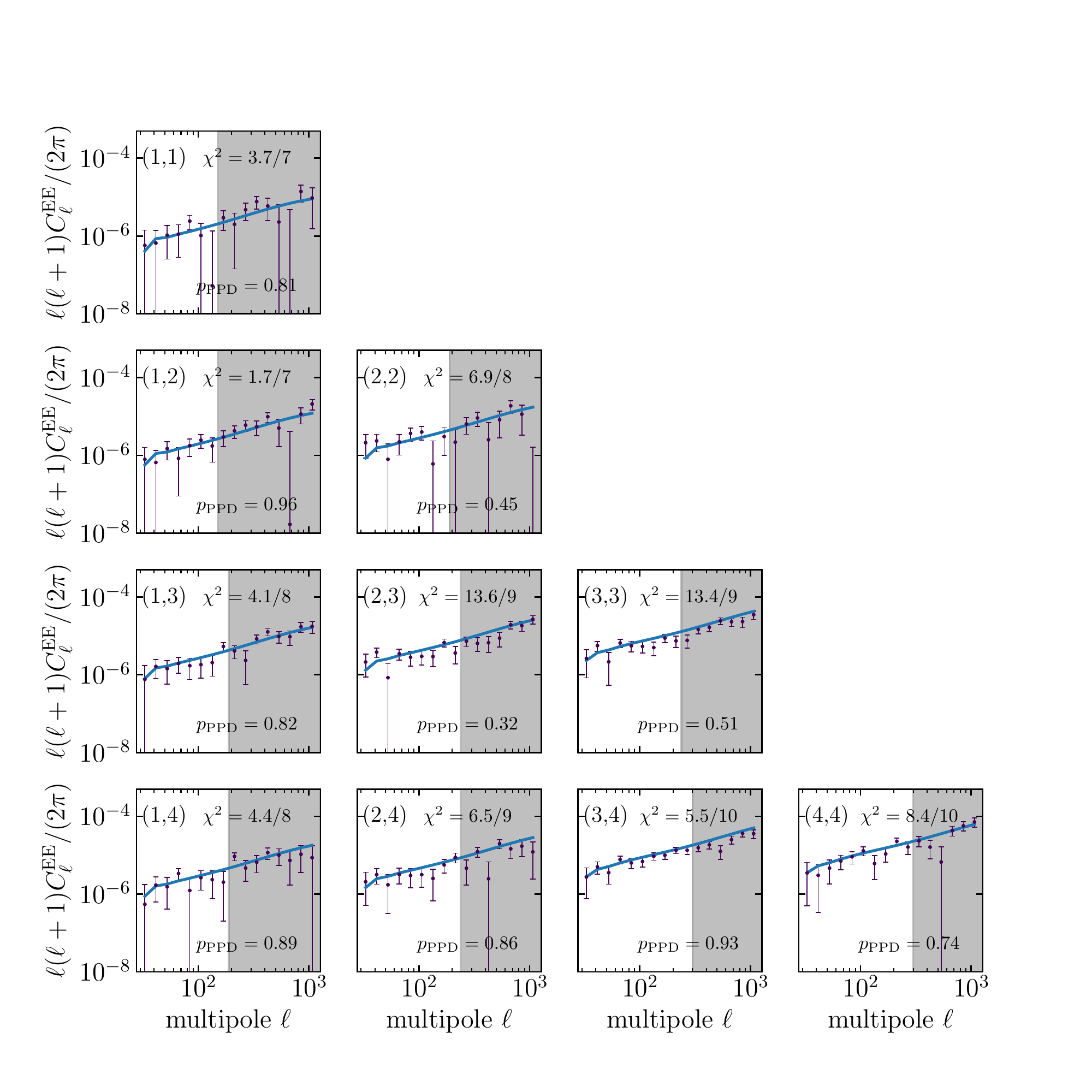}
  \caption{The cosmic shear angular power spectra for the \mcal catalog.
    Error-bars are the diagonal elements of the fiducial covariance matrix.
    The continuous line shows the recovered best-fit model. The vertical
    shaded region shows the scale-cuts applied. After considering the
    scale-cuts, the recovered $\chi^2$ obtained is 65.5 for 69 degrees of
    freedom.}
  \label{fig:bf_metacal}
\end{figure*}

\begin{figure*}
  \centering
  \includegraphics[width=0.99\textwidth]{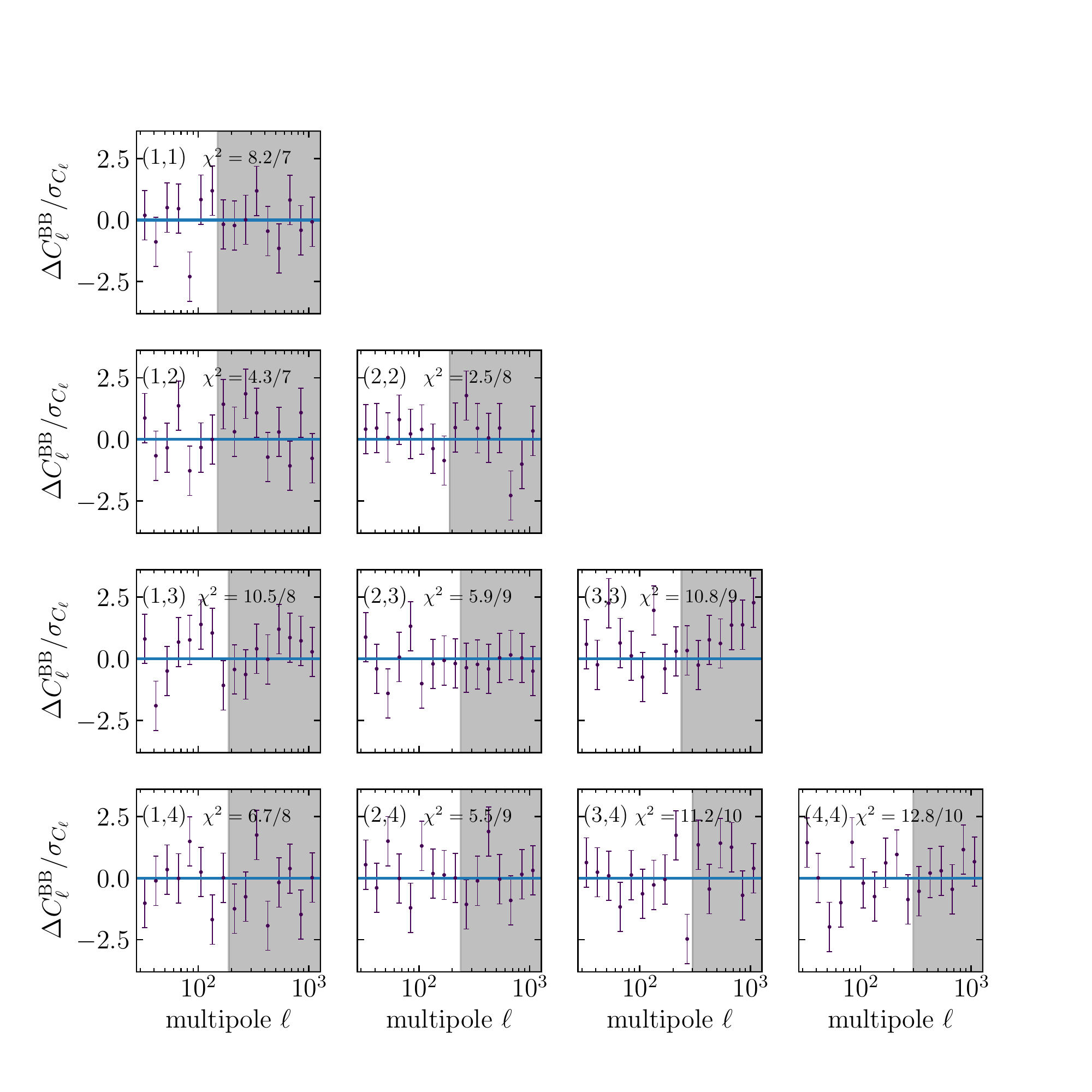}
  \caption{The cosmic-shear $B$-modes angular power spectra for the \mcal
    catalog. Error bars are the diagonal elements of the fiducial covariance
    matrix. The reference is the null model after bandpower binning. The
    vertical shaded region shows the scale-cuts applied. The goodness of fit
    for each spectrum is shown on each panel. Combining all the spectra into a
    single data vector yields a $\chi^2 = 78.3$.}
  \label{fig:BB}
\end{figure*}

\begin{figure*}
  \centering
  \includegraphics[width=0.99\textwidth]{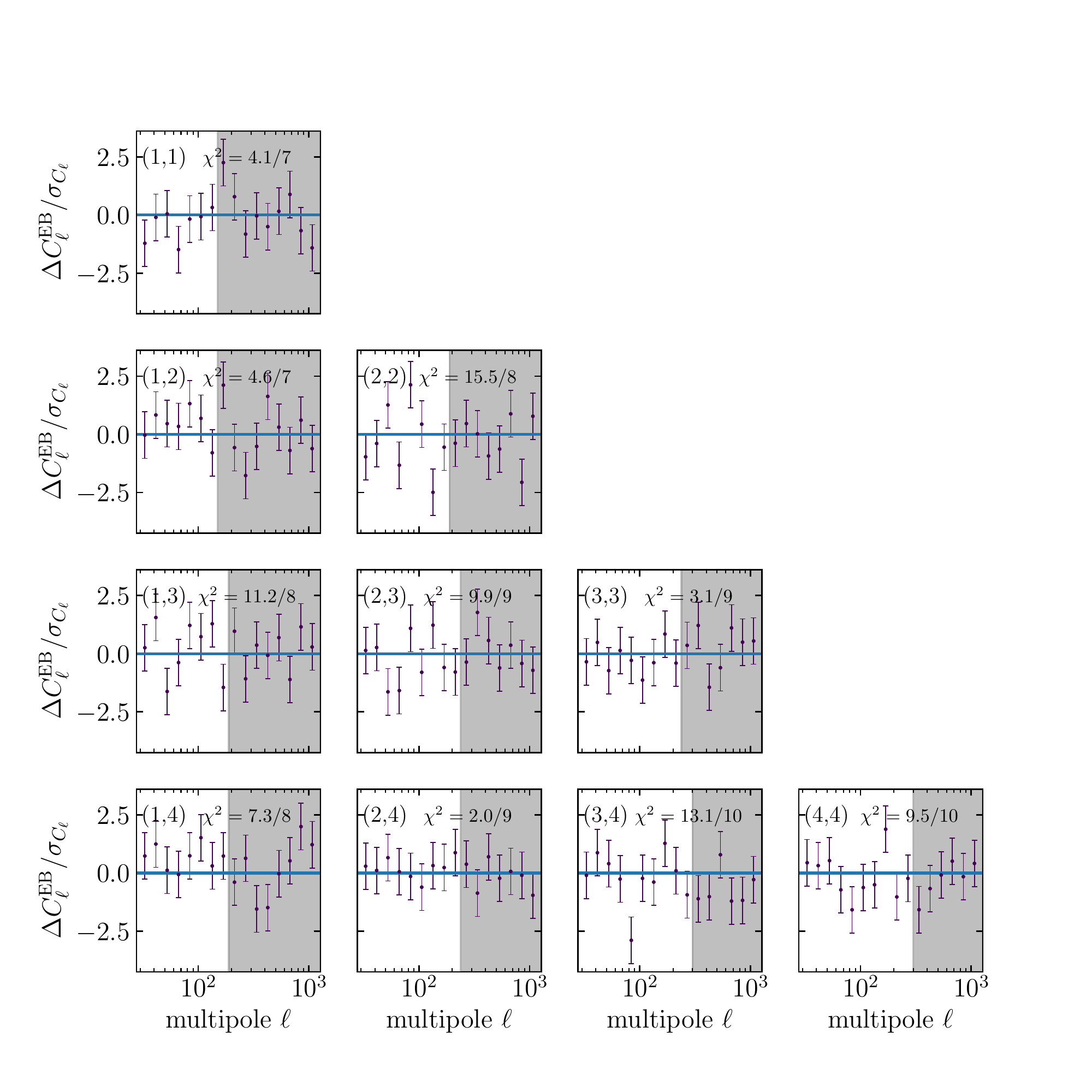}
  \caption{Same as Figure \ref{fig:BB} but for the $EB$-modes angular power
    spectra measured on the \mcal catalog. Combining all the spectra into a
    single data vector yields a $\chi^2 = 80.3$.}
  \label{fig:EB}
\end{figure*}

We have developed a pipeline for Bayesian parameter inference, constructed by
adapting the existing DES-Y1 \cosmosis pipeline developed for a
configuration-space analysis \citep{krause:2016jvl} to perform an analysis in
harmonic space. We also use this existing pipeline for all our results quoting
configuration space. We use the nested sampling technique for the sampling of
Markov Chain Monte Carlo (MCMC) chains. In particular, we use the publicly
available \multinest code\footnote{The \multinest configuration parameters used
for the analysis were: \texttt{live\_points}=501, \texttt{efficiency}=0.3,
\texttt{tolerance}=0.1 and \texttt{constant\_efficiency}=F.}
\citep{Feroz_2009,2019OJAp....2E..10F}. We used a Gaussian likelihood, $L$, defined as
\begin{equation}
  -2 \log L (\Theta) = \chi^2 = \sum_{ij} (\hat{D}_i - D_i(\Theta))^{T}
  \mathrm{C}^{-1}_{ij} (\hat{D}_j - D_{j} (\Theta)),
  \label{eqn:chi2}
\end{equation}
where $\hat{D}_i$ are the entries of the data vector, constructed by stacking the measured power spectra bandpowers $C^{EE(ab)}(\ell)$ for the different combinations of tomographic bins pairs, $(a, b)$, accounting for scale-cuts, see Sections \ref{sec:Measurements} and \ref{sec:ScaleCuts} and Table \ref{tab:scale-cuts} and $D_i(\Theta)$ are their theoretical predictions computed according to the modelling presented in Section \ref{sec:formalism}.
Finally, $\Theta$ represents the set of parameters, cosmological and nuisance, used in the analysis, see Table \ref{tab:Parameters}, and $\mathrm{C}$ is the covariance matrix, see Section \ref{sec:MethodsCovariance}.

\subsection{Validation with a noiseless data vector}

The first step in the analysis is to validate our pipeline 
using a noiseless analytically computed data vector generated with the \flask cosmology. 
We used our pipeline for the
likelihood analyses both in configuration and harmonic space, with and without
scale cuts motivated to mitigate baryonic effects and the same binning as
described in Section~\ref{sec:ScaleCuts}. 
This data vector does not contain baryonic effects since the scale cuts were chosen in such a way that they become unimportant, see Section~\ref{sec:ScaleCuts}.
Therefore we do not expect baryonic effects to be relevant in our test with the adopted scale cuts, which is why we employed a noiseless data vector. 
Our aim in this subsection is to test the consistency of the pipeline.
For the configuration space run, we
follow the DES-Y1 setup \citep{troxel_2018} again.
Figure~\ref{fig:corner_hsvscs} shows the 2D posterior probability distributions
and constraints for a subset of the inferred parameters, namely $S_8$,
$\Omega_m$ and $\sigma_8$. Figure~\ref{fig:robustness_flask} provides the 1D
marginalized values of the parameters $S_8$, $\Omega_m$, $\sigma_8$, $h_0$,
$\Omega_b$ and $n_s$. We conclude that the likelihood pipeline is working as
expected in this case, with consistent values for the recovered parameters and
error bars.

\subsection{Cosmic shear likelihood analysis in DES-Y1}

\begin{table}
  \caption{The marginalised constraints for the fiducial analysis on
    configuration and harmonic space. We quote the mean value of the
    marginalised posterior distribution and the 68\% confidence level (CL)
    around it, as well as the associated $\chi^2$. After applying scale-cuts,
    the data vectors have 227 and 85 elements for the configuration and
    harmonic space case, respectively. We have 16 model parameters for both
    cases, yielding 211 and 69 d.o.f.\ for the CS and HS cases, respectively.
    These constraints are also presented as error bars in Figure
    \ref{fig:robustness}. }
  \label{tab:hsvscs_metacal}
  \begin{tabular*}{\columnwidth}{@{}l@{\hspace*{20pt}}c@{\hspace*{20pt}}c@{\hspace*{20pt}}c@{}}
    \hline
    Case & $\chi^2/\mathrm{d.o.f.}$ & $\Omega_m$ & $S_8$ \\
    \hline
    \textbf{HS, Updated cov} & \textbf{65.5/69} &
    $\mathbf{0.304_{-0.042}^{+0.067}}$ & $\mathbf{0.766\pm 0.033}$ \\ HS,
    Fiducial cov & 62.8/69 & $0.302^{+0.042}_{-0.073}$ &
    $0.765_{-0.036}^{+0.032}$ \\ CS & 230.0/211 & $0.295^{+0.040}_{-0.059}$ &
    $0.778_{-0.029}^{+0.024}$ \\ \hline HS, Gaussian cov & 65.6 / 69 &
    $0.305_{-0.038}^{+0.077}$ & $0.767\pm 0.034$ \\ HS, FLASK cov & 53 / 69 &
    $0.300_{-0.035}^{+0.066}$ & $0.765^{+0.036}_{-0.033}$ \\ HS,
    $\ell_\mathrm{min}=38$ & 60.5 / 59 & $0.287^{+0.035}_{-0.065}$ &
    $0.764_{-0.038}^{+0.033}$ \\ HS, Fixed $\Omega_{\nu}$ & 65.34 / 70 &
    $0.298_{-0.038}^{+0.066}$ & $0.764\pm 0.034$ \\ HS, No IA & 66.7 / 71 &
    $0.305_{-0.038}^{+0.077}$ & $0.767\pm 0.034$ \\
    \hline
  \end{tabular*}
\end{table}

We now proceed to the likelihood analysis of the DES-Y1 data.
The estimated power spectra for DES-Y1 data are shown in fig. \ref{fig:bf_metacal}, along with the recovered best-fit model for our fiducial $\Lambda$CDM results.
We begin with a couple of null test validations on the data. 
First, in the Born approximation,
cosmological shear should not produce B-modes. However, in practice, they can
be generated by the masking procedure. In \cite{zuntz:2017pso} it was already
shown that the \mcal catalogue does not contain significant contamination by
$B$-modes. Here we extend this tests and verify that the procedure of
recovering the true $C_{\ell}^{EE}$ does not introduce significant
contributions to $C_{\ell}^{BB}$ in Figure~\ref{fig:BB} and also
$C_{\ell}^{EB}$ in Figure~\ref{fig:EB}. The figure presents the residuals of
the measurements with respect to a null model, $\Delta C_\ell^{BB/EB}$
normalised by the standard deviation extracted from the fiducial covariance
matrix. The measured $C_{\ell}^{BB}$ and $C_{\ell}^{EB}$ are consistent with a
null angular power spectrum after the binning procedure with a reasonable
$\chi^2$ per degree of freedom. 
We recall here that, to properly account for the binning of the null spectra model in the Pseudo-$C_\ell$ estimation context, we follow \cite{2019MNRAS.484.4127A} and apply the bandpower window function, as in Equation \eqref{eq:bpws-modeling}.

\begin{figure*}
  \centering
  \includegraphics[width=0.99\textwidth]{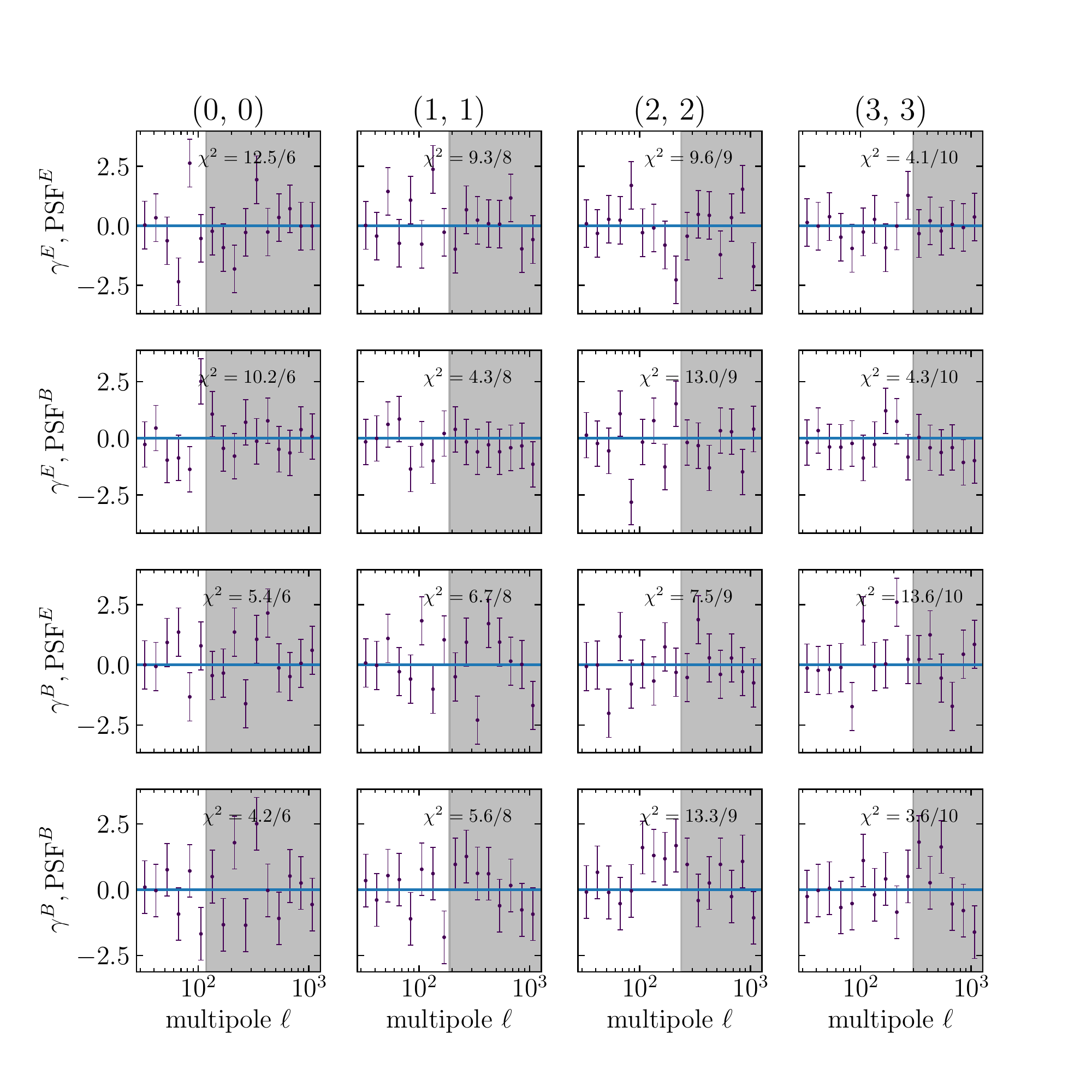}
  \caption{The cosmic-shear correlations between signal and PSF for the \mcal
    catalogue. Error bars are the diagonal elements of the fiducial covariance
    matrix. The reference is the null model after bandpower binning. The
    vertical shaded region shows the scale-cuts applied. The goodness of fit
    for each spectrum is shown on each panel.
    }
  \label{fig:psf}
\end{figure*}

Secondly, it is well known that the point spread function (PSF) distorts the
images of the galaxies and if not modelled properly, it can lead to significant
systematic errors. In order to check its impact on our measurements, we use
PSF maps estimated for the DES-Y1 \mcal catalogue \citep{zuntz:2017pso} to
estimate its correlation with the $E/B$-mode of the shear signal,
$\gamma^{E/B}$. The result is presented in Figure~\ref{fig:psf}, where each
column presents the 4 different combinations of the PSF $E/B$ maps and shear
$E/B$ maps for a tomographic bin $a \in \{0, 1, 2, 3\}$. As for the previous
null test, we summarize the results presenting the residuals with respect to a
null signal model normalized by the standard deviation from the fiducial
covariance. Our results suggest consistency of these cross-correlations with a
null signal. Therefore, we do not apply any further systematic correction on
the measured shear spectra.

We then focus on the extraction of cosmological information from the measured
$C_\ell^{EE}$ power spectra. We vary the six cosmological parameters and the
ten nuisance parameters with fiducial values and priors shown in
Table~\ref{tab:Parameters}
\footnote{It has been claimed that the DES Y1 priors on $\Omega_m$ and $\sigma_8$ may suffer from small prior volume effects \cite{2021A&A...646A.129J}.
However, this effect is not important for constraints on $S_8$.}.
Neutrino masses were varied using three degenerate
neutrinos, following \cite{troxel_2018}. The nuisance parameters that enter
the theoretical modelling of the systematic effects are marginalised to extract
cosmological information. We also run the DES-Y1 shear analysis in
configuration space to compare the cosmological constraining power of both
analyses.

\begin{figure}
  \centering
  \includegraphics[width=0.49\textwidth]{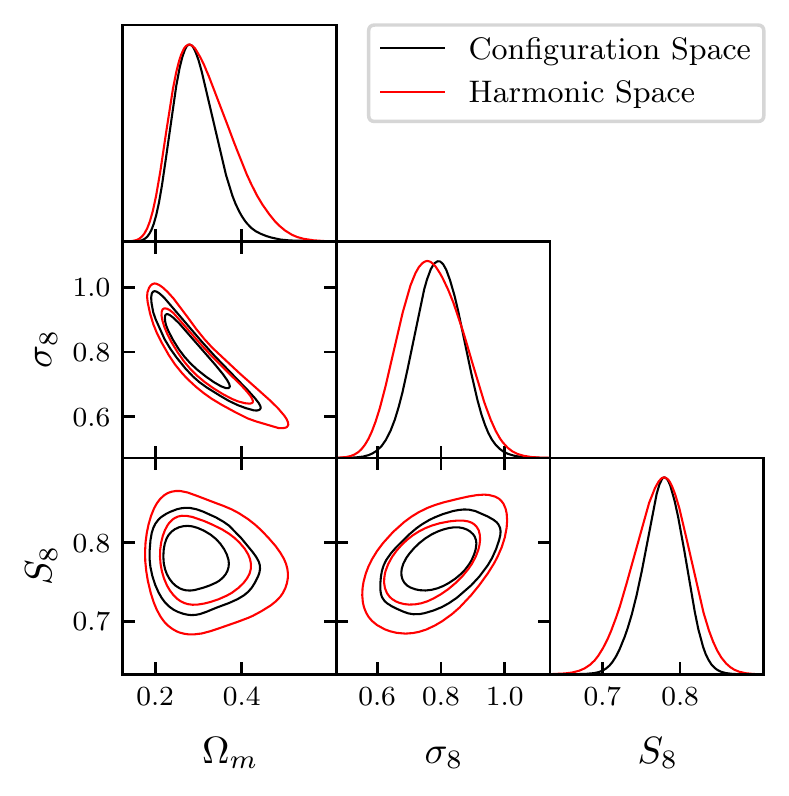}
  \caption{Marginalized posterior distributions for a subset of constrained
    parameters. We show the results for configuration- and harmonic-space, see
    Table~\ref{tab:hsvscs_metacal}. }
  \label{fig:corner_hsvscs_metacat_marg}
\end{figure}

\begin{figure}
  \centering
  \includegraphics[width=0.49\textwidth]{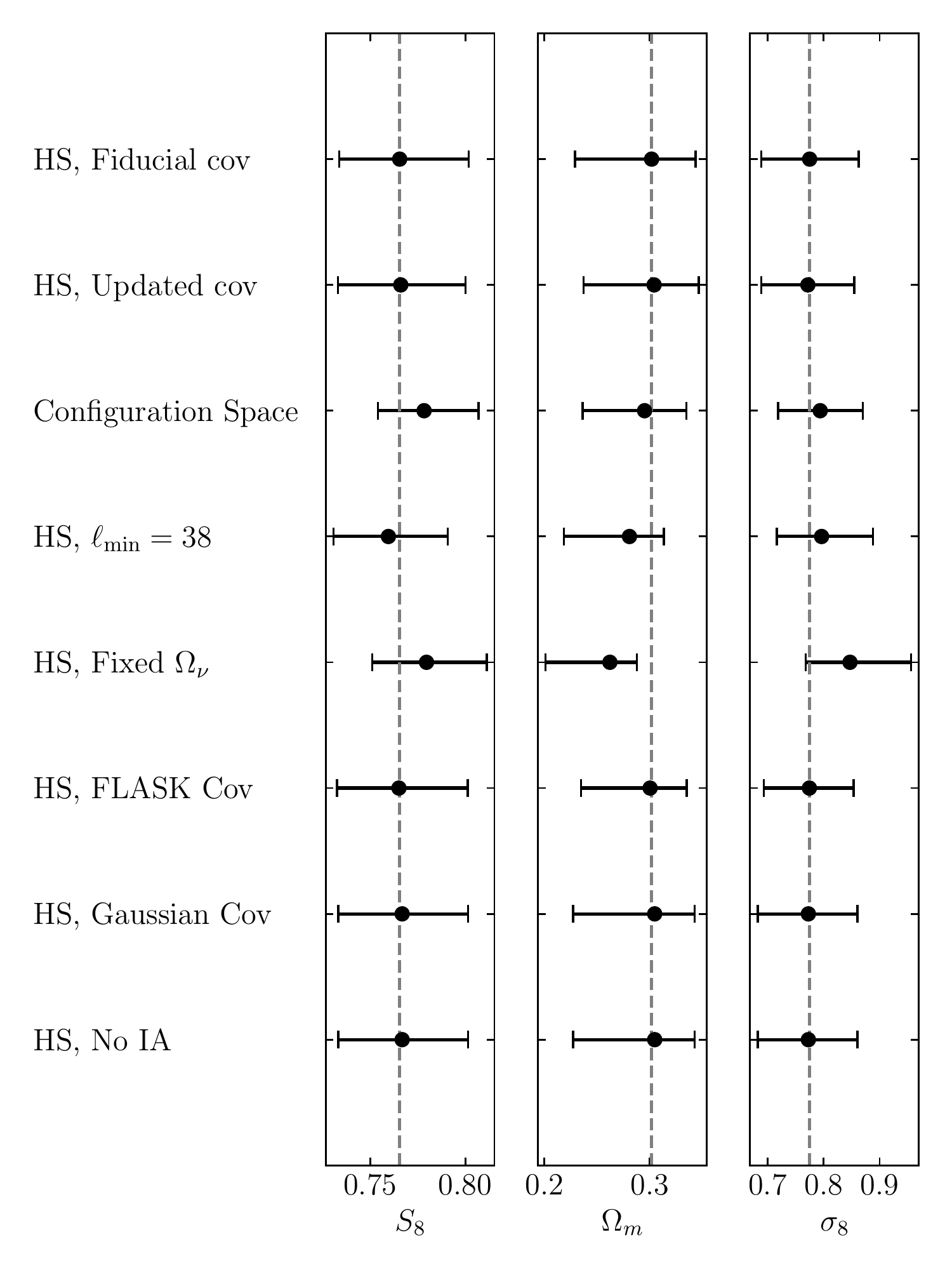}
  \caption{The summary of the one dimensional marginalised constraints on
    $S_8$, $\Omega_m$ and $\sigma_8$. The 68\% CL are shown as error bars
    around the mean value for the recovered posterior PDF presented as the
    central point. A battery of robustness tests presented as variations to
    the fiducial analysis setup are presented. We further present a set of
    robustness tests representing variations to the fiducial analysis setup are
    presented. The associated numerical values are presented in
    Table~\ref{tab:hsvscs_metacal}. }
  \label{fig:robustness}
\end{figure}

Finally, we re-run the whole harmonic space analysis with an updated covariance
matrix, with the Gaussian part computed at the cosmological parameters obtained
from the best fit. Our main results are shown in
Figures~\ref{fig:corner_hsvscs_metacat_marg}, \ref{fig:robustness} and
Table~\ref{tab:hsvscs_metacal} for the 2-D and 1-D marginalized posterior
probability distribution on the main cosmological parameters $\Omega_m$, $S_8$
and $\sigma_8$ from a likelihood analysis in both configuration and harmonic
space.

We find very good agreement between the two different analyses. The errors are
comparable and cosmological parameters are in agreement within less than one
standard deviation for both parameter, more precisely, less than $\sim 0.2 \,
\sigma$ for $\Omega_m$ and less than $\sim 0.4 \,\sigma$ for $S_8$. The
$\chi^2$ per degree of freedom are consistent and demonstrate a good
quality-of-fit for both analyses. The quality-of-fit for each pair of bins are
also shown in Figure~\ref{fig:bf_metacal}. We also present an additional test
on the posterior predictive distribution (PPD), following the methodology
presented in \citet{2021MNRAS.503.2688D}. Namely, the PPD goodness-of-fit
test, the probability-to-exceed quantified by the $p$-value, $p_{\rm PPD}$, is
also displayed for each pair of bins considered.

On top of well consistent constraints, we found variations of $<1\%$ in the
$\chi^2$ when consider a pure Gaussian covariance matrix, suggesting a
negligible impact for the non-Gaussian corrections in our analysis. This is
consistent with results in configuration space \citep{troxel_2018} and the
analysis in harmonic space presented by \cite{Nicola:2020lhi}. It is important
to note that the latter reports variations of $\Delta\chi^2 \sim 1\%$ but
$<10\%$. The lower differences founded here can be understood as a result of
our treatment of baryonic effects and the resulting scale cuts.

\section{Robustness tests}
\label{sec:robustness}

In this section, we perform a number of robustness tests of our analysis in
harmonic space.

\begin{itemize}

\item \textbf{Impact of the covariance cosmology.}  Our analysis was performed
  with a theoretical covariance matrix computed at the \flask cosmology. In
  this subsection, we update the covariance matrix to the best-fit cosmological
  parameters of our analysis in harmonic space and re-run our likelihood
  pipeline. The results are shown in Figure \ref{fig:robustness}, and there
  are no significant changes with respect to the original covariance matrix.

  In addition, we also studied the changes in the estimated cosmological
  parameters arising from using the estimated sample covariance from the suite
  of lognormal \flask realisations. When using this sample covariance, our
  approach is to use a Gaussian likelihood correcting only the covariance by
  the Hartlap-Anderson factor \citep{2007A&A...464..399H}. We tested the
  effect of changing the likelihood to a $t$-student function as motivated by
  \cite{2016MNRAS.456L.132S} founding no appreciable differences. As can be
  seen in Table \ref{tab:hsvscs_metacal} again no significant changes are
  found.

\item \textbf{Scale cuts.}  Data from large scales are affected by the geometry
  effects of the mask. These effects are in principle dealt with using the
  filtering prescription of \citet{2019MNRAS.484.4127A} that we adopt here. In
  this subsection, we test the large scale cuts used in the fiducial choice by
  leaving out the first $\ell$ bin, using $\ell_\mathrm{min}=38$ instead of
  $\ell_\mathrm{min}=30$. As seen in Table \ref{tab:hsvscs_metacal} again no
  significant changes are found.

\end{itemize}


\section{Conclusions}
\label{sec:conclusions}

We have presented a cosmological analysis using the cosmic shear angular power
spectrum obtained from measurements of the DES-Y1 \mcal shear catalogue. We
closely follow the configuration space shear 2-point analysis of
\citet{troxel_2018}, including the theoretical modelling of redshift
uncertainties, shear calibration and intrinsic alignments.

We validated our pipeline using a suite of 1200 lognormal \flask mocks. The
analysis choices and scale cuts were imposed following a similar prescription
for baryon contamination as the configuration space analysis. Our analytical
covariance matrix was obtained from combining a Gaussian contribution that
incorporates the survey geometry in the so-called improved Narrow-Kernel
Approximation (iNKA) approximation with a non-Gaussian contribution from
\cosmolike and validated using shear measurements of the 1200 \flask mocks.
The shape noise contributions to the power spectra and the covariance matrices were estimated analytically following \cite{Nicola:2020lhi} using the so-called sum of weights mask scheme, see Section \ref{sec:Measurements}.
We used our pipeline to measure the angular power spectrum $C_{\ell}^{EE}$ in the
DES-Y1 \mcal catalogue and show that it does not introduce significant
contributions to $C_{\ell}^{BB}$ and $C_{\ell}^{EB}$.

Finally, we performed a likelihood analysis using the \cosmosis framework in
both configuration space, reproducing the DES-Y1 results, and in harmonic space
with a fiducial analysis and also study the impact of variations, such as a
covariance matrix computed in a different cosmology, a scale cut on large
scales, and a different treatment of neutrino masses. Although the analysis in
configuration and harmonic space are independent, we find results for the
cosmological parameters $S_8$ and $\Omega_m$ that are very consistent.
Differences were found to be less than $\sim 0.2 \, \sigma$ for $\Omega_m$ and
less than $\sim 0.4 \,\sigma$ for $S_8$. These results are encouraging and
provide a stepping stone to the shear analysis in harmonic space using the
third year of DES data (Y3). The DES-Y3 shear analysis in harmonic space will
use a similar pipeline but with some improved modelling, mostly following the
methodology laid out for the real-space case \citep{krause2021} and the
real-space shear results \citep{secco2021,amon2021}: an inverse-variance
weights determined in the Y3 \mcal catalogue \citep{gatti2021}, a tidal
alignment and tidal torquing (TATT) model \citep{TATT} for intrinsic alignment,
a determination of scale cuts using a $\chi^2$ criterion between noiseless data
vectors with and without baryon effect contamination, the usage of a blinding
strategy and more robustness tests. The results will be addressed in a
forthcoming publication.


\section*{Acknowledgements}

This research was partially supported by the Laborat\'orio Interinstitucional
de e-Astronomia (LIneA), the Brazilian funding agencies CNPq and CAPES, the
Instituto Nacional de Ci\^{e}ncia e Tecnologia (INCT) e-Universe (CNPq grant
465376/2014-2) and the Sao Paulo State Research Agency (FAPESP) through grants
2019/04881-8 (HC) and 2017/05549-1 (AT). The authors acknowledge the use of
computational resources from LIneA, the Center for Scientific Computing
(NCC/GridUNESP) of the Sao Paulo State University (UNESP), and from the
National Laboratory for Scientific Computing (LNCC/MCTI, Brazil), where the
SDumont supercomputer ({\tt sdumont.lncc.br}) was used. This research used
resources of the National Energy Research Scientific Computing Center (NERSC),
a U.S. Department of Energy Office of Science User Facility operated under
Contract No. DE-AC02-05CH11231.

This paper has gone through internal review by the DES collaboration. Funding
for the DES Projects has been provided by the U.S. Department of Energy, the
U.S. National Science Foundation, the Ministry of Science and Education of
Spain, the Science and Technology Facilities Council of the United Kingdom, the
Higher Education Funding Council for England, the National Center for
Supercomputing Applications at the University of Illinois at Urbana-Champaign,
the Kavli Institute of Cosmological Physics at the University of Chicago, the
Center for Cosmology and Astro-Particle Physics at the Ohio State University,
the Mitchell Institute for Fundamental Physics and Astronomy at Texas A\&M
University, Financiadora de Estudos e Projetos, Funda{\c c}{\~a}o Carlos Chagas
Filho de Amparo {\`a} Pesquisa do Estado do Rio de Janeiro, Conselho Nacional
de Desenvolvimento Cient{\'i}fico e Tecnol{\'o}gico and the Minist{\'e}rio da
Ci{\^e}ncia, Tecnologia e Inova{\c c}{\~a}o, the Deutsche
Forschungsgemeinschaft and the Collaborating Institutions in the Dark Energy
Survey.

The Collaborating Institutions are Argonne National Laboratory, the University
of California at Santa Cruz, the University of Cambridge, Centro de
Investigaciones Energ{\'e}ticas, Medioambientales y Tecnol{\'o}gicas-Madrid,
the University of Chicago, University College London, the DES-Brazil
Consortium, the University of Edinburgh, the Eidgen{\"o}ssische Technische
Hochschule (ETH) Z{\"u}rich, Fermi National Accelerator Laboratory, the
University of Illinois at Urbana-Champaign, the Institut de Ci{\`e}ncies de
l'Espai (IEEC/CSIC), the Institut de F{\'i}sica d'Altes Energies, Lawrence
Berkeley National Laboratory, the Ludwig-Maximilians Universit{\"a}t
M{\"u}nchen and the associated Excellence Cluster Universe, the University of
Michigan, the National Optical Astronomy Observatory, the University of
Nottingham, The Ohio State University, the University of Pennsylvania, the
University of Portsmouth, SLAC National Accelerator Laboratory, Stanford
University, the University of Sussex, Texas A\&M University, and the OzDES
Membership Consortium.

Based in part on observations at Cerro Tololo Inter-American Observatory at
NSF's NOIRLab (NOIRLab Prop. ID 2012B-0001; PI: J. Frieman), which is managed
by the Association of Universities for Research in Astronomy (AURA) under a
cooperative agreement with the National Science Foundation.

The DES data management system is supported by the National Science Foundation
under Grant Numbers AST-1138766 and AST-1536171. The DES participants from
Spanish institutions are partially supported by MINECO under grants
AYA2015-71825, ESP2015-66861, FPA2015-68048, SEV-2016-0588, SEV-2016-0597, and
MDM-2015-0509, some of which include ERDF funds from the European Union. IFAE
is partially funded by the CERCA program of the Generalitat de Catalunya.
Research leading to these results has received funding from the European
Research Council under the European Union's Seventh Framework Program
(FP7/2007-2013) including ERC grant agreements 240672, 291329, and 306478.

This manuscript has been authored by Fermi Research Alliance, LLC under
Contract No. DE-AC02-07CH11359 with the U.S. Department of Energy, Office of
Science, Office of High Energy Physics.

This work made use of the software packages {\tt matplotlib}
\citep{matplotlib}, and {\tt numpy} \citep{numpy}.


\section*{Data Availability Statement}

The DES Y1 catalog is available in the Dark Energy Survey Data Management
(DESDM) system at the National Center for Supercomputing Applications (NCSA) at
the University of Illinois. It can be accessed at
\url{https://des.ncsa.illinois.edu/releases/y1a1/key-catalogs}. The pipeline
used for the measurement is publicly available at
\url{https://github.com/hocamachoc/3x2hs_measurements}. Synthetic data produced
by the analysis presented here can be shared on request to the corresponding
author.


\bibliographystyle{mnras}
\bibliography{literature}

\section*{Author Affiliations}
\label{app:affiliations}
$^{1}$ Instituto de F\'{i}sica Te\'orica, Universidade Estadual Paulista, S\~ao Paulo, Brazil\\
$^{2}$ Laborat\'orio Interinstitucional de e-Astronomia - LIneA, Rua Gal. Jos\'e Cristino 77, Rio de Janeiro, RJ - 20921-400, Brazil\\
$^{3}$ ICTP South American Institute for Fundamental Research\\ Instituto de F\'{\i}sica Te\'orica, Universidade Estadual Paulista, S\~ao Paulo, Brazil\\
$^{4}$ Departamento de F\'isica Matem\'atica, Instituto de F\'isica, Universidade de S\~ao Paulo, CP 66318, S\~ao Paulo, SP, 05314-970, Brazil\\
$^{5}$ Department of Physics and Astronomy, University of Pennsylvania, Philadelphia, PA 19104, USA\\
$^{6}$ Department of Astronomy, University of California, Berkeley,  501 Campbell Hall, Berkeley, CA 94720, USA\\
$^{7}$ Department of Astronomy/Steward Observatory, University of Arizona, 933 North Cherry Avenue, Tucson, AZ 85721-0065, USA\\
$^{8}$ Jet Propulsion Laboratory, California Institute of Technology, 4800 Oak Grove Dr., Pasadena, CA 91109, USA\\
$^{9}$ Kavli Institute for Cosmology, University of Cambridge, Madingley Road, Cambridge CB3 0HA, UK\\
$^{10}$ Department of Physics, Northeastern University, Boston, MA 02115, USA\\
$^{11}$ Laboratory of Astrophysics, \'Ecole Polytechnique F\'ed\'erale de Lausanne (EPFL), Observatoire de Sauverny, 1290 Versoix, Switzerland\\
$^{12}$ Jodrell Bank Center for Astrophysics, School of Physics and Astronomy, University of Manchester, Oxford Road, Manchester, M13 9PL, UK\\
$^{13}$ California Institute of Technology, 1200 East California Blvd, MC 249-17, Pasadena, CA 91125, USA\\
$^{14}$ Kavli Institute for Particle Astrophysics \& Cosmology, P. O. Box 2450, Stanford University, Stanford, CA 94305, USA\\
$^{15}$ Lawrence Berkeley National Laboratory, 1 Cyclotron Road, Berkeley, CA 94720, USA\\
$^{16}$ Institut d'Estudis Espacials de Catalunya (IEEC), 08034 Barcelona, Spain\\
$^{17}$ Institute of Space Sciences (ICE, CSIC),  Campus UAB, Carrer de Can Magrans, s/n,  08193 Barcelona, Spain\\
$^{18}$ Faculty of Physics, Ludwig-Maximilians-Universit\"at, Scheinerstr. 1, 81679 Munich, Germany\\
$^{19}$ Department of Astronomy, University of Geneva, ch. d'\'Ecogia 16, CH-1290 Versoix, Switzerland\\
$^{20}$ Department of Applied Mathematics and Theoretical Physics, University of Cambridge, Cambridge CB3 0WA, UK\\
$^{21}$ Department of Astronomy and Astrophysics, University of Chicago, Chicago, IL 60637, USA\\
$^{22}$ Kavli Institute for Cosmological Physics, University of Chicago, Chicago, IL 60637, USA\\
$^{23}$ Department of Physics, Carnegie Mellon University, Pittsburgh, Pennsylvania 15312, USA\\
$^{24}$ Brookhaven National Laboratory, Bldg 510, Upton, NY 11973, USA\\
$^{25}$ Department of Physics, Duke University Durham, NC 27708, USA\\
$^{26}$ Institut de F\'{\i}sica d'Altes Energies (IFAE), The Barcelona Institute of Science and Technology, Campus UAB, 08193 Bellaterra (Barcelona) Spain\\
$^{27}$ Institute for Astronomy, University of Edinburgh, Edinburgh EH9 3HJ, UK\\
$^{28}$ Cerro Tololo Inter-American Observatory, NSF's National Optical-Infrared Astronomy Research Laboratory, Casilla 603, La Serena, Chile\\
$^{29}$ Fermi National Accelerator Laboratory, P. O. Box 500, Batavia, IL 60510, USA\\
$^{30}$ Institute of Cosmology and Gravitation, University of Portsmouth, Portsmouth, PO1 3FX, UK\\
$^{31}$ CNRS, UMR 7095, Institut d'Astrophysique de Paris, F-75014, Paris, France\\
$^{32}$ Sorbonne Universit\'es, UPMC Univ Paris 06, UMR 7095, Institut d'Astrophysique de Paris, F-75014, Paris, France\\
$^{33}$ Department of Physics \& Astronomy, University College London, Gower Street, London, WC1E 6BT, UK\\
$^{34}$ SLAC National Accelerator Laboratory, Menlo Park, CA 94025, USA\\
$^{35}$ Center for Astrophysical Surveys, National Center for Supercomputing Applications, 1205 West Clark St., Urbana, IL 61801, USA\\
$^{36}$ Department of Astronomy, University of Illinois at Urbana-Champaign, 1002 W. Green Street, Urbana, IL 61801, USA\\
$^{37}$ Physics Department, William Jewell College, Liberty, MO, 64068, USA\\
$^{38}$ Astronomy Unit, Department of Physics, University of Trieste, via Tiepolo 11, I-34131 Trieste, Italy\\
$^{39}$ INAF-Osservatorio Astronomico di Trieste, via G. B. Tiepolo 11, I-34143 Trieste, Italy\\
$^{40}$ Institute for Fundamental Physics of the Universe, Via Beirut 2, 34014 Trieste, Italy\\
$^{41}$ Observat\'orio Nacional, Rua Gal. Jos\'e Cristino 77, Rio de Janeiro, RJ - 20921-400, Brazil\\
$^{42}$ Department of Physics, University of Michigan, Ann Arbor, MI 48109, USA\\
$^{43}$ Hamburger Sternwarte, Universit\"{a}t Hamburg, Gojenbergsweg 112, 21029 Hamburg, Germany\\
$^{44}$ Centro de Investigaciones Energ\'eticas, Medioambientales y Tecnol\'ogicas (CIEMAT), Madrid, Spain\\
$^{45}$ Department of Physics, IIT Hyderabad, Kandi, Telangana 502285, India\\
$^{46}$ Santa Cruz Institute for Particle Physics, Santa Cruz, CA 95064, USA\\
$^{47}$ Department of Astronomy, University of Michigan, Ann Arbor, MI 48109, USA\\
$^{48}$ Institute of Theoretical Astrophysics, University of Oslo. P.O. Box 1029 Blindern, NO-0315 Oslo, Norway\\
$^{49}$ Instituto de Fisica Teorica UAM/CSIC, Universidad Autonoma de Madrid, 28049 Madrid, Spain\\
$^{50}$ School of Mathematics and Physics, University of Queensland,  Brisbane, QLD 4072, Australia\\
$^{51}$ Center for Cosmology and Astro-Particle Physics, The Ohio State University, Columbus, OH 43210, USA\\
$^{52}$ Department of Physics, The Ohio State University, Columbus, OH 43210, USA\\
$^{53}$ Center for Astrophysics $\vert$ Harvard \& Smithsonian, 60 Garden Street, Cambridge, MA 02138, USA\\
$^{54}$ Australian Astronomical Optics, Macquarie University, North Ryde, NSW 2113, Australia\\
$^{55}$ Lowell Observatory, 1400 Mars Hill Rd, Flagstaff, AZ 86001, USA\\
$^{56}$ George P. and Cynthia Woods Mitchell Institute for Fundamental Physics and Astronomy, and Department of Physics and Astronomy, Texas A\&M University, College Station, TX 77843,  USA\\
$^{57}$ Department of Astrophysical Sciences, Princeton University, Peyton Hall, Princeton, NJ 08544, USA\\
$^{58}$ Instituci\'o Catalana de Recerca i Estudis Avan\c{c}ats, E-08010 Barcelona, Spain\\
$^{59}$ Physics Department, 2320 Chamberlin Hall, University of Wisconsin-Madison, 1150 University Avenue Madison, WI  53706-1390\\
$^{60}$ Institute of Astronomy, University of Cambridge, Madingley Road, Cambridge CB3 0HA, UK\\
$^{61}$ School of Physics and Astronomy, University of Southampton,  Southampton, SO17 1BJ, UK\\
$^{62}$ Computer Science and Mathematics Division, Oak Ridge National Laboratory, Oak Ridge, TN 37831\\
$^{63}$ Department of Physics, Stanford University, 382 Via Pueblo Mall, Stanford, CA 94305, USA\\
$^{64}$ Max Planck Institute for Extraterrestrial Physics, Giessenbachstrasse, 85748 Garching, Germany\\
$^{65}$ Universit\"ats-Sternwarte, Fakult\"at f\"ur Physik, Ludwig-Maximilians Universit\"at M\"unchen, Scheinerstr. 1, 81679 M\"unchen, Germany\\
$^{66}$ Department of Physics and Astronomy, Pevensey Building, University of Sussex, Brighton, BN1 9QH, UK\\

\appendix

\section{Residual systematics in the cosmic shear signal}
\label{sec:residual-systematics}

\begin{figure*}
    \centering
    \includegraphics{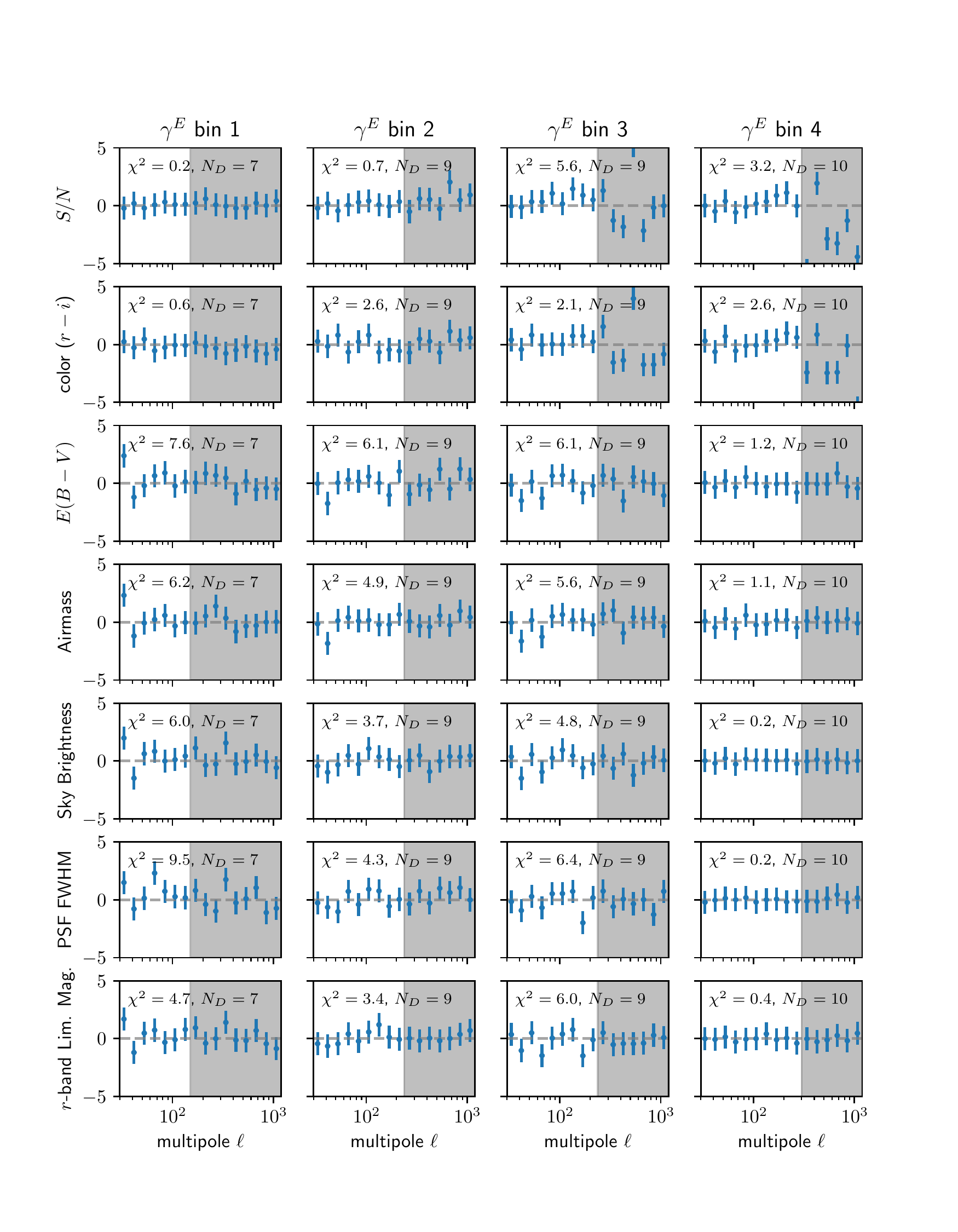}
    \caption{Testing for residual systematics in the tomographic cosmic shear signal in harmonic space
    Each panel shows measured cross-correlation of the different tomographic bins considered in this work (columns) and a subsample of survey properties most likely to be sourcing residual shear systematics (rows).
    The fiducial analysis used the same bandpower binning and scale cuts (shaded regions). 
    The significance of the null-test, $\chi^2$, and the number of elements in the data vector are shown in each panel.}
    \label{fig:residual-systematics}
\end{figure*}

The DES Y1 \mcal catalogue has been carefully designed for testing for systematics under a battery of null tests \citep{zuntz:2017pso}, resulting in the advice of accounting for possible photo-z and shear estimation systematic biases as done in the present work.
However, potential residual systematics biases that have not been identified can persist.
Following the DES SV \citep{2016PhRvD..94b2002B} and DES Y1 \citep{troxel_2018} cosmic shear analyses in configuration space, we test those by considering a subsample of survey properties that are most likely to be sourcing residual shear systematics. 
On top of the PSF ellipticity presented in \ref{sec:robustness}, we consider signal-to-noise ($S/N$), $r-i$ color, dust extinction ($E(B-V )$), sky brightness, PSF size (PSF FWHM), airmass, and $r$-band limiting magnitude.
The first four are intrinsic properties of each galaxy image measured by the DES Y1 shape and PSF measurement pipelines \citep{zuntz:2017pso}.
The last five are the mean value of each property across exposures at a given position in the sky. 
We generated \healpix maps for those properties with the same \healpix resolution of our measurements, NSIDE of 1024.
As in \citep{troxel_2018}, we do not consider several properties tested in the DES SV analysis because of their high degeneracy with the considered ones. Also, as catalogue preparation for DES Y1 data found no need to make an explicit surface brightness cut in the shape catalogues \citep{zuntz:2017pso}, we do not consider that property.

Our methodology is, however, different from the one from configuration space analyses. 
We consider the cross-correlation between the observed shear signal and the survey properties and test for a null hypothesis quantified in the $\chi^2$ using the fiducial covariance and scale-cuts of the analysis.

We present our result in the Figure \ref{fig:residual-systematics}, where the estimated cross-correlations normalised by the error bars are presented.
The figure also quotes the significance, $\chi^2$, and the number of points in the considered data vector, $N_D$.
There is no strong evidence of cross correlation between survey properties and the shear signal in any of the tomographic bins.

For the intrinsic properties of each galaxy image, $S/N$ and color, there does seem to be higher significance for cross-correlation in the highest redshift bin for the smallest scales, cut out by the scale-cuts in our analysis. For the rest of the properties, our first bandpower, $[30, 37)$, exhibits the most considerable significance of cross-correlation, not statistically significant when combined with the rest of the data vector for any of the cases.

\section{Full marginalized 2D and 1D posteriors}

\begin{figure*}
    \centering
    \includegraphics[width=\textwidth]{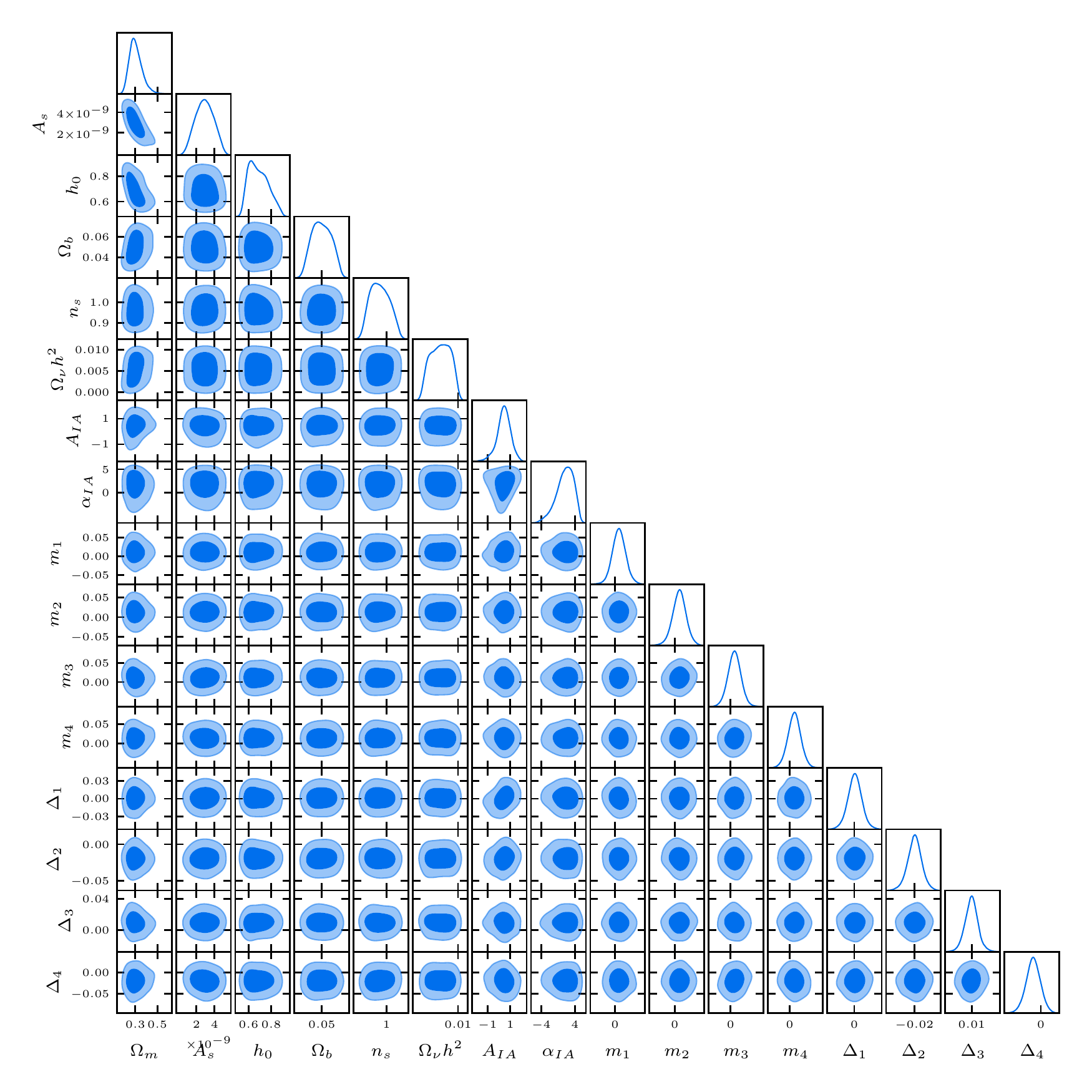}
    \caption{The full marginalized 2D and 1D posterior PDF for all 16 parameters in our fiducial $\Lambda$CDM model.
    The 2D contours show the 68\% and 95\% confidence intervals.}
    \label{fig:full2d-posterior}
\end{figure*}

We show all the marginalised 2D and 1D posteriors for the full parameter space of our fiducial $\Lambda$CDM analysis in Figure \ref{fig:full2d-posterior}.
No significant constraint beyond the prior was found for all the nuisance parameters, $m_{i}, \Delta z_i$, nor for $h_0, \Omega_b, n_s, \Omega_\nu h^2$.

\bsp
\label{lastpage}
\end{document}